\documentclass[11pt]{article}

\newcommand{\rr}{\mathbb R}

\newcommand{\inv}{^{-1}}
\newcommand{\abs}[1]{\left|{#1}\right|}

\newcommand{\suchthat}{\ | \ }

\newcommand{\tth}{^\text{th}}
\newcommand{\genseq}[3]{{#1}_1 {#3} {#1}_2 {#3} \dots {#3} {#1}_{#2}}
\newcommand{\seq}[2]{\genseq{#1}{#2}{,}}

\newcommand{\threecases}[6]{\begin{cases} #2 & #1 \\ #4 & #3 \\ #6 & #5 \end{cases}}
\newcommand{\txt}[1]{\text{#1}}
\newcommand{\stext}[1]{\ \ \ \ \ \text{(#1)}}
\newcommand{\stextn}[1]{\\&\ \ \ \ \ \ \stext{#1}}

\newcommand{\snc}[1]{\stext{since ${#1}$}}
\newcommand{\push}{\\ & \ \ \ \ \ \ \ \ \ \ }
\newcommand{\nrm}[1]{\mathcal{N}\left(#1\right)}
\newcommand{\ee}{\mathop{\mathbb{E}}}

\newcommand{\ipnc}[3]{\begin{figure}[ht]\begin{center}\includegraphics[scale = {#1}]{#2.pdf}\caption{#3}\end{center}\end{figure}}
\usepackage[linesnumbered,algoruled,boxed,lined]{algorithm2e}
\makeatletter 
\g@addto@macro{\@algocf@init}{\SetKwInOut{Parameter}{Parameters}} 
\makeatother

\newcommand{\lp}[3]{\begin{tabular}{l l}\textbf{#1} & \begin{tabular}{l}$#2$\end{tabular}\\\textbf{subject to} & \begin{tabular}{l l}#3\end{tabular}\end{tabular}}
\newcommand{\lpmax}[2]{\lp{maximize}{#1}{#2}}

\usepackage{amsmath}
\usepackage{amssymb}
\usepackage{graphicx}
\usepackage{float}
\usepackage{amsthm}
\usepackage{color}
\usepackage{enumitem}
\usepackage[caption=false]{subfig}
\usepackage{authblk}

\theoremstyle{plain}
\newtheorem{theorem}{Theorem}
\newtheorem{lemma}[theorem]{Lemma} 
\newtheorem{proposition}[theorem]{Proposition}

\numberwithin{theorem}{section}

\title{Playing Divide-and-Choose Given Uncertain Preferences\footnote{This is a pre-print of our publication in \emph{Management Science}. The published version is available here:\\\texttt{https://pubsonline.informs.org/doi/10.1287/mnsc.2023.00350}}}
\author[1]{Jamie Tucker-Foltz}
\author[2]{Richard Zeckhauser}
\affil[1]{Harvard University, \texttt{jtuckerfoltz@gmail.com}}
\affil[2]{Harvard University, \texttt{richard\_zeckhauser@harvard.edu}}

\begin{document}
\maketitle
\begin{abstract}
	We study the classic divide-and-choose method for equitably allocating divisible goods between two players who are rational, self-interested Bayesian agents. The players have additive values for the goods. The prior distributions on those values are common knowledge. We consider both the cases of independent values and values that are correlated across players (as occurs when there is a common-value component).
	
	We describe the structure of optimal divisions in the divide-and-choose game and identify several cases where it is possible to efficiently compute equilibria. An approximation algorithm is presented for the case when the distribution over the chooser's value for each good follows a normal distribution, along with a randomized approximation algorithm for the case of uniform distributions over intervals.
	
	A mixture of analytic results and computational simulations illuminates several striking differences between optimal strategies in the cases of known versus unknown preferences. Most notably, given unknown preferences, the divider has a compelling ``diversification'' incentive in creating the chooser's two options. This incentive leads to multiple goods being divided at equilibrium, quite contrary to the divider's optimal strategy when preferences are known.
	
	In many contexts, such as buy-and-sell provisions between partners, or in judging fairness, it is important to assess the relative expected utilities of the divider and chooser. Those utilities, we show, depend on the players' levels of knowledge about each other's values, the correlations between the players' values, and the number of goods being divided. Under fairly mild assumptions, we show that the chooser is strictly better off for a small number of goods, while the divider is strictly better off for a large number of goods.
\end{abstract}

\section{Introduction}\label{secIntro}

Ever since Abraham and Lot divided the land of Canaan, with Abraham dividing and Lot choosing, the divide-and-choose method has been employed to parcel out assets. Today the method is widely used when partners in a real estate deal invoke their buy-sell agreement, or when siblings divide up an inheritance.

The assets can be continuous, as with the classic cake-cutting problem, or separable and indivisible, for example valuables from an estate. Consider multi-dimensional cake-cutting with divisible elements vanilla filling, chocolate icing and a cherry. One player will divide the cake; the other will then choose her piece. Positing additive preferences and that fractional portions have fractional values, if the divider knows the chooser's preferences, he arrays the assets (filling, icing, cherry) in order of the ratio of his value to her (the chooser's) value. Just enough low-ratio assets are placed in pile 2 to assure that the chooser selects it, leaving the divider the high-ratio assets in pile 1.\footnote{Throughout this paper, we assume without loss of generality that ``pile 1'' is the divider's preferred pile.} Only one asset will generally need to be divided fractionally; in knife-edge situations it could be none.

Now consider a more consequential example: Partners in a real-estate company have signed a buy-sell arrangement. As is common, after a specified period of years, either partner can trigger the arrangement by creating a two-pile division. Pile 1 might consist of asset $Q$ and a required payment of \$1.5 million. Pile 2 would contain the remaining asset, $R$, and receipt of \$1.5 million. Partner A launches as the divider. Partner B must then choose between the two piles. The very nature of the legal agreement allows A to reject any proposals from B to adjust terms.

In real life, the players rarely know each others' preferences. The cake divider will know his own preferences, but will only have a feel---that is, a Bayesian prior---for how the chooser values filling to icing to cherry, or asset $Q$ to asset $R$ to money. While the divider can think in terms of expected ratios, he will remain uncertain of the chooser's total value for any two piles. The divider's optimal strategy will weigh the disadvantages of putting fewer goods (or lesser amounts of divisible goods) in pile 1 against the likelihood that the chooser picks pile 1. How that trade-off should be handled is the subject of this paper.

\subsection{Our contributions}\label{subContributions}

Our central contribution is to analyze the divide-and-choose game when information is asymmetric, i.e.\txt{}, when the players’ values are private information, but priors for those values are common knowledge. This is the typical setup for self-interested players in Bayesian games. Section \ref{secModel} presents our model for divide-and-choose.

The literature on divide-and-choose games focuses overwhelmingly on the case of known preferences, though real-world situations rarely meet that standard. In the comforting land of known preferences, the divider follows the simple optimization procedure mentioned above. In sharp contrast, once uncertainty enters, a strategic Bayesian divider generally faces a complicated optimization problem, potentially with a myriad of local optima. In Section \ref{secComputation} we show that such complexities can arise even for extremely simple priors, such as independently and identically distributed (i.i.d.) normal distributions.

The apparent lack of structure to the divider's optimization problem suggests that it may be computationally intractable in general. One of our main contributions is an algorithmic framework for optimizing over divisions that yields polynomial-time approximation algorithms when the distributions over the chooser's values for each good satisfy certain properties. Specifically, when each distribution is symmetric about its mean and log-concave, we show that there exists a sequence of convex programs computing divisions that yield expected utility arbitrarily close to the maximum possible expected utility (Theorem \ref{thmSymmetricConvexity}). As special cases, we obtain a fully polynomial-time approximation scheme (FPTAS) when the distributions are Gaussian (Theorem \ref{thmNormalAlgoFPTAS}) and a fully polynomial-time \emph{randomized} approximation scheme (FPRAS) when the distributions are uniform over real intervals (Theorem \ref{thmUniformOpt}). We also present an algorithm that solves the case of discrete distributions, which is practical when the number of possible chooser types is small.

This complexity of the divider's optimization problem arises because, in contrast to the case of known preferences, it is often beneficial to divide more than one good. To take our cake analogy, if a divider prefers filling and suspects that the chooser has a high value for both the icing and the cherry, pile 1 may optimally consist of 100\% of the filling, plus 59\% of the icing and 31\% of the cherry. By ensuring that pile 2 contains a bit of icing and cherry, he decreases the variance in the chooser’s valuation for pile 1, hence reducing the likelihood that she will choose that pile. In Section \ref{secDiversification} we investigate the divider's incentive to diversify his risk by dividing several goods (or, for indivisible assets, committing to lotteries that randomly allocate some goods between the piles after the chooser picks a pile). Even for a risk-neutral divider, diversification is warranted for a wide range of prior distributions. This is still the case even when the divider's values are strongly correlated with those of the chooser. Furthermore, we show that a risk averse divider should diversify more. These results still do not determine \emph{which} goods should be split between the two piles. Simple rules, we show, can lead the divider astray.

Finally, in Section \ref{secWelfare} we analyze the expected utilities of the divider and chooser under a range of circumstances. For example, which player is better off \emph{a priori} if both players' values are drawn from the same distribution?  With known preferences, it is far preferable to be the divider. With significant uncertainties about the players' values of the individual goods, the chooser is better off when there are few goods, but the advantage tips to the divider as the number of goods increases.

Our Conclusion, Section \ref{secConclusion}, presents a number of open questions that emerged from this analysis. It closes by stressing that although our focus has been on theory and computational methods, the analysis is widely applicable in real-world contexts. The divide-and-choose method, or close analogues, though rarely identified that way, are widely used in practice. Take-it-or-leave-it offers and buy-and-sell agreements between business partners are salient examples. 

\subsection{Related work}\label{subLitReview}

The divide-and-choose method features prominently in the literature on \emph{fair division}. In the \emph{cake-cutting} model \cite{CakeCuttingSurvey}, agents are assumed to have additive, divisible preferences over the unit interval $[0, 1]$ (the ``cake'') and a feasible allocation is a partitioning of $[0, 1]$ between the players, where each player's part of the allocation is a finite union of intervals. With two players, the divide-and-choose method is particularly useful in this model for the following reasons:
\begin{enumerate}[label={(\roman*)}]
	\item If the first player divides the cake equally, the allocation will be \emph{envy-free}, meaning that the players each value their own piece of cake higher.
	\item The protocol can be implemented by asking the players simple queries.
\end{enumerate}
Extensions have also been discovered for 3 or more players \cite{BramsTaylor, EFUpperBound}. These methods require several rounds of dividing and choosing, but ultimately satisfy the same two properties. Kuhn \cite{kuhn1967games} gives abstract axioms under which a simple one-round variant of divide-and-choose with one divider and $n - 1$ choosers will yield an acceptable outcome for all $n$ players. Another major line of work studies deterministic allocations of indivisible goods, in which case envy-freeness is sometimes not feasible. Think of three goods each valued roughly the same by both players. They can at best be divided 1 and 2, so the divider will generally be envious. Hence, various natural relaxations have been studied \cite{EF1, Prop1}. Variant settings with more complex kinds of goods and constraints have also been studied, such as goods with unknown, random values \cite{BMS}.

The fair division literature has largely focused on finding mechanisms satisfying axiomatic properties such as envy-freeness, efficiency, etc., when players do \emph{not} behave strategically. A handful of works consider the objective of strategy-proofness as well, both for divisible \cite{DivisibleGoodsMechDesign1,DivisibleGoodsMechDesign2,DivisibleGoodsMechDesign4,DivisibleGoodsMechDesign5,TruthJusticeCake} and indivisible \cite{IndivisibleGoodsMechDesign1,IndivisibleGoodsMechDesign2} goods. These works largely draw on the mechanism design perspective, studying questions such as, what is the optimal worst-case utility approximation to the socially optimal outcome under the constraint of strategy-proofness?

Our work is more in the spirit of Clausewitz the strategist than Solomon the arbitrator: Our prime interest is not in finding mechanisms to implement socially desirable outcomes. Rather, we focus on how strategic players should actually behave to maximize their personal welfare. Issues of welfare and efficiency enter our analysis, particularly how those issues are affected by some simple extensions of the divide-and-choose game. However, our goal is not to modify the game to bolster its fairness or efficiency, but rather to analyze the effects from a descriptive perspective. There is a small experimental literature on strategic behavior under various cake-cutting protocols; for instance, see Kyropoulou, Ortega, and Segal-Halevi \cite{FairCCInPractice}. On the theoretical front, however, much less is known. We are aware of one prior work, by Delgosha and Gohari \cite{FCCIPRef1}, that studies optimal strategies in an environment where players may learn about each other's preferences through repeated interactions. Our interest is in the more common setting of one-shot interactions.

A key concept we discuss is the \emph{critical ratio} of a good, which we define to be the ratio of the divider's value to the expectation of the chooser's value. This is a ubiquitous notion in contexts from hypothesis testing to cost/effectiveness analyses. (See, for example, Weinstein and Zeckhauser \cite{CriticalRatios}.) In the case where the divider has perfect knowledge of the chooser's preferences, the divider places the goods with the highest critical ratio in pile 1 and those with the lowest critical ratio in pile 2. He determines the cutoff so that the chooser just picks pile 2. The optimality of this strategy in the setting of cake-cutting is noted by H.\txt{} P.\txt{} Young \cite{HPYoungBook}, as well as Br\^anzei, Caragiannis, Kurokawa, and Procaccia \cite{AlgorithmicFramework}. One of the key conceptual takeaways from this paper is that uncertainty over preferences notably complicates the task of computing optimal divisions in a way that is not characterized by critical ratios.

Finally, we note that our model bears a technical resemblance to a model of final-offer arbitration (FOA) by Powers \cite{Arbitration} and Brams and Merrill \cite{ArbitrationOlder}. Just as in our setting, there are multiple divisible goods that must be allocated among two players. However, instead of playing divide-and-choose, both players simultaneously submit divisions to an arbiter, who chooses the more ``reasonable'' one, where players have common knowledge of the arbiter's sense of reasonableness. A key technical difference between FOA and divide-and-choose is that, in FOA, the uncertainty is not over the other player's values, but over the other player's action. Yet both models involve similar optimal strategies: give just enough away to the other player to obtain the desired binary decision with decent probability.

\section{Model}\label{secModel}

There are two players: the \emph{divider} ($D$) and the \emph{chooser} ($C$). There are $n$ divisible goods, which we number from 1 to $n$, writing $[n] := \{1, 2, 3, \dots, n\}$ for the set of goods. We may also think of the goods as indivisible, in which case we allow the divider to fractionally allocate goods between the piles via lotteries that are resolved after the chooser has chosen her pile. A player's value is simply the sum of the values of the goods received. Players are risk-neutral, except in Section \ref{subDiversificationRiskAversion} on risk aversion. For each $i \in [n]$, we denote the respective private values of good $i$ to the divider and chooser by $g^D_i$ and $g^C_i$, which are drawn from a joint distribution $\mathcal{G}_i$ over $\rr^2$. We write $\mathcal{G}^D_i$ and $\mathcal{G}^C_i$ for the respective marginal distributions. More compactly, we denote the value vectors $g^D := (g^D_1, g^D_2, \dots, g^D_n)$ and $g^C := (g^C_1, g^C_2, \dots, g^C_n)$, which are drawn according to the joint distribution $\mathcal{G} = \mathcal{G}_1 \times \mathcal{G}_2 \times \dots \times \mathcal{G}_n$. Thus, we allow for values of the two players for any individual good to be correlated, but require independence across goods. We write the respective marginal distributions $\mathcal{G}^D := \mathcal{G}^D_1 \times \mathcal{G}^D_2 \times \dots \times \mathcal{G}^D_n$ and $\mathcal{G}^D := \mathcal{G}^C_1 \times \mathcal{G}^C_2 \times \dots \times \mathcal{G}^C_n$. We write $\overline{\mathcal{G}}^C_i$ and $\overline{\mathcal{G}}^C$ for the updated distributions over the chooser's values after the divider observes his values. The divide-and-choose game proceeds in three steps.
\begin{enumerate}[label={(\roman*)}]
	\item\label{itmDCGameObserve} Values are drawn $(g^D, g^C) \sim \mathcal{G}$. The divider observes $g^D$ and the chooser observes $g^C$.
	\item\label{itmDCGameDivide} The divider chooses a \emph{division} of the goods, which is a vector $p = (\seq{p}{n}) \in [0, 1]^n$. We refer to \emph{pile 1} as the allocation consisting of $p_i$ of each good $i$ and \emph{pile 2} as the allocation consisting of $(1 - p_i)$ of each good $i$.
	\item\label{itmDCGameChoose} The chooser picks her higher-valued pile, namely pile 1 or pile 2, for herself. The other pile goes to the divider.
\end{enumerate}
Formally, the payoffs are defined as follows. If the chooser picks pile 1, then the divider receives a payoff of
$u^D = \sum_{i = 1}^{n} (1 - p_i) g^D_i$
and the chooser receives a payoff of
$u^C = \sum_{i = 1}^{n} p_i g^C_i.$
On the other hand, if the chooser picks pile 2, then the divider receives a payoff of
$u^D = \sum_{i = 1}^{n} p_i g^D_i$
and the chooser receives a payoff of
$u^C = \sum_{i = 1}^{n} (1 - p_i) g^C_i.$
To facilitate understanding, we refer to pile 1 as the pile that the divider would prefer to have for himself, an assumption that is without loss of generality (see Lemma \ref{lemStructureGeneral}). Given a division $p$, we refer to the probability that the chooser picks pile 1 as $P$. 

Thus, the game is completely parameterized by the value distribution $\mathcal{G}$. In this paper we mostly focus on settings where the values are independent across players, except in Sections \ref{subDiversificationCorrelation} and \ref{subWelfareDividerVersusChooser}. We especially focus on the following distributions:
\begin{itemize}
	\item \emph{Normal priors}: The value for good each $i$ is drawn from $\mathcal{N}(\mu_i, \sigma_i)$ for some mean $\mu_i \in \rr$ and standard deviation $\sigma_i > 0$.
	\item \emph{Discrete priors}: The value for each good $i$ is drawn from an arbitrary distribution over $\rr$ with finite support.
	\item \emph{Uniform priors}: The value for each good $i$ is drawn i.i.d.\txt{} from uniform distributions over intervals $[a_i, b_i]$, where $0 \leq a_i \leq b_i$.
\end{itemize}

Throughout this paper we often assume that each $g^D_i$ has been fixed, which makes the distribution $\mathcal{G}^D_i$ irrelevant. Given fixed values of $g^D_i$, we define the \emph{critical ratio} of good $i$ to be
$$CR_i := \frac{g^D_i}{\ee[g^C_i]},$$
where the expectation is over $\overline{\mathcal{G}}_i$, i.e., the conditional expectation of $g^C_i$ given $g^D_i$.

Note that our model allows for goods to take potentially negative values (i.e., be ``bads''). With normal priors this will happen with some nonzero probability, though in most of our examples the means are sufficiently large to render this probability negligibly small. For some of our results we also require that values be nonnegative. For instance, we only discuss critical ratios in the context where all divider values and chooser means are positive.

Our solution concept is that of a subgame-perfect, Bayes-Nash equilibrium. Without loss of generality we need only consider pure strategies for both players, since at no point can either player benefit from inferring information about the other player's type from its actions. We make a minor additional technical assumption: indifference is always broken in favor of the other player. This is only necessary for knife-edge cases in some of our theorems. In practice, we would expect optimal divisions and optimal pile choices to be unique, rendering this assumption unnecessary.

\section{Computing optimal divisions}\label{secComputation}

The divider first observes his values for the $n$ goods $g^D_1, g^D_2, \dots, g^D_n$. How should he use that information together with the updated priors\footnote{Throughout this section we use the word ``prior'' to refer to the distribution of chooser values conditioned on the divider's values.} $\overline{\mathcal{G}}^C_1, \overline{\mathcal{G}}^C_2, \dots, \overline{\mathcal{G}}^C_n$ on the chooser's values to determine his optimal division? We first establish some basic, general properties of the equilibria of the divide-and-choose game. We then describe algorithms to determine the divider's optimal division for both normal priors and discrete priors. Finally, we generalize our algorithm for normal priors to handle a wider class of distributions, one that includes uniform priors.

\subsection{The general structure of optimal divisions}

We begin by recording two easy facts about optimal divisions, which we will use throughout the paper. Their straightforward proofs are deferred to Appendices \ref{appBaselineUtilities} and \ref{appProofStructureGeneral}.

The first observation, to use terminology from the fair division literature, says that all equilibria satisfy \emph{proportionality} for the chooser and \emph{expected proportionality} for the divider. In other words, both players can expect to take away at least half of their total values.

\begin{lemma}\label{lemBaselineUtilities}
	In any equilibrium of the divide-and-choose game, the following hold.
	\begin{enumerate}[label={(\roman*)}]
		\item\label{itmBaselineChooser} For any realizations of $g^D$ and $g^C$,
		$u^C \geq \frac12 \sum_{i = 1}^n g^C_i.$
		\item\label{itmBaselineDivider} For any realization of $g^D$,
		$\ee_{g^C \sim \overline{\mathcal{G}}^C}[u^D] \geq \frac12 \sum_{i = 1}^n g^D_i.$
	\end{enumerate}
\end{lemma}

We refer to the utilities
$\frac12 \sum_{i = 1}^n g^D_i \ \txt{ and }\ \frac12 \sum_{i = 1}^n g^C_i$
as the respective \emph{proportionality guarantees} utilities of the divider and chooser.

It is convenient to consider the divider's optimization problem, not employing the $p_i$ variables taking values in $[0, 1]$, but instead in terms of the auxiliary variables
\begin{equation}\label{equDefineQFromP}
	q_i := 2p_i - 1 = p_i - (1 - p_i),
\end{equation}
taking values in $[-1, 1]$. For reference, the inverse of this correspondence is
\begin{equation}\label{equDefinePFromQ}
	p_i = \frac{q_i}{2} + \frac12.
\end{equation}
We frequently refer to a division as $p$ or $q$ interchangeably.

\begin{lemma}\label{lemStructureGeneral}
	A divider-optimal division with the following two properties always exists.
	\begin{itemize}
		\item The divider weakly prefers pile 1:
		\begin{equation}\label{equDividerPrefersPile1Q}
			\sum_{i = 1}^{n} q_i g^D_i \geq 0.
		\end{equation}
		\item The chooser is weakly more likely to pick pile 2:
		\begin{equation}\label{equChooserProbablyPrefersPile2Q}
			P := \Pr_{g^C \sim \overline{\mathcal{G}}^C}\left[\sum_{i = 1}^n q_ig^C_i \geq 0\right] \leq \frac12.
		\end{equation}
	\end{itemize}
	Furthermore, the divider can achieve a strictly higher interim expected utility than his proportionality guarantee if and only if both of these inequalities can be made strict.
\end{lemma}

For a wide class of priors over the chooser's value, including normal and uniform distributions, we can characterize exactly when the inequalities from Lemma \ref{lemStructureGeneral} will be strict.

\begin{proposition}\label{proStructureNormal}
	Suppose each $\overline{\mathcal{G}}^C_i$ is a non-atomic distribution that is symmetric about its mean, and suppose all divider values and chooser means are positive. Then the optimal division yields utility equal to the proportionality guarantee if and only if all goods have the same critical ratios.
\end{proposition}

\newcommand{\contentProofStructureNormal}{
	Suppose first that all goods have the same critical ratio $r$. First note that the division $q_1 = q_2 = \dots = q_n = 0$ For any other division $q$, where at least one $q_i$ is not zero, consider the random variable
	\begin{equation}\label{equDefineX}
		X := \sum_{i = 1}^n q_i g^C_i.
	\end{equation}
	Since the distribution of $X$ is a nontrivial linear combination of the $\overline{\mathcal{G}}^C_i$, it is also non-atomic and symmetric about its mean, which implies that, for any threshold $t$,
	\begin{equation}\label{equStructureNormalProperty}
		\Pr[X > t] > \frac12 \iff \ee[X] > T.
	\end{equation}
	Therefore,
	\begin{align*}
		P > \frac12 \iff \Pr\left[\sum_{i = 1}^n q_i g^C_i > 0\right] > \frac12
		&\iff \ee\left[\sum_{i = 1}^n q_i g^C_i\right] > 0 \stext{from (\ref{equStructureNormalProperty})}\\
		\iff \ee\left[\sum_{i = 1}^n q_i g^D_i \frac{g^C_i}{g^D_i}\right] > 0
		&\iff \sum_{i = 1}^n q_i g^D_i \frac{\ee\left[g^C_i\right]}{g^D_i} > 0\\
		&\iff r\sum_{i = 1}^n q_i g^D_i > 0\\
		&\iff \sum_{i = 1}^n q_i g^D_i > 0 \stext{since $r$ is positive by assumption}.
	\end{align*}
	In other words, the chooser is strictly more likely to pick pile 1 if and only if the divider strictly prefers pile 1. That it turn implies that it is not possible to achieve a higher-than-proportional utility by the final statement of Lemma \ref{lemStructureGeneral}.
	
	Now suppose that two goods $j$ and $k$ have different critical ratios, i.e., $\overline{\mathcal{G}}^C_j$ has mean $\mu_j$ and $\overline{\mathcal{G}}^C_k$ has mean $\mu_k$ such that,
	$$\frac{g^D_j}{\mu_j} > \frac{g^D_k}{\mu_k}.$$
	Let $\alpha := \max\{g^D_j + \mu_j, g^D_k + \mu_k\}$, let $q$ be the division such that
	\begin{align*}
		q_j &:= \frac{g^D_k + \mu_k}{\alpha}\\
		q_k &:= -\frac{g^D_j + \mu_j}{\alpha}\\
		q_i &:= 0 \stext{for all $i \in [n] \setminus \{j, k\}$}.
	\end{align*}
	Note that $q_i = 0$ corresponds to $p_i = \frac12$, so in words, this is a division where good $j$ is slightly more in pile 1, good $k$ is slightly more in pile 2, and all other goods are divided equally between the two piles (scaling by $\alpha$ ensures $\abs{q_j}, \abs{q_k} \leq 1$). Then
	$$\sum_{i = 1}^n q_i g^D_i = q_j g^D_j + q_k g^D_k = \frac{g^D_k + \mu_k}{\alpha} g^D_j - \frac{g^D_j + \mu_j}{\alpha} g^D_k = \frac{\mu_k g^D_j - \mu_j g^D_k}{\alpha} > 0,$$
	so the divider strictly prefers pile 1. Also, the random variable $X$ has mean
	$$\sum_{i = 1}^n q_i \mu_i = q_j \mu_j + q_k \mu_k = \frac{g^D_k + \mu_k}{\alpha} \mu_j - \frac{g^D_j + \mu_j}{\alpha} \mu_k = \frac{g^D_k \mu_j - g^D_j \mu_k}{\alpha} < 0,$$
	which implies (via (\ref{equStructureNormalProperty})) that
	$$P = \Pr\left[\sum_{i = 1}^n q_i g^C_i > 0\right] < \frac12.$$
	Therefore, it follows from the final statement of Lemma \ref{lemStructureGeneral} that the divider achieves a utility that is higher than his proportionality guarantee.
}

\begin{proof}
	\contentProofStructureNormal
\end{proof}

We remark that this result does not extend to some other natural families of distributions, for instance, discrete 2-point distributions. In fact, even when all chooser priors and divider values are identical, there may exist bizarre ``symmetry-breaking'' divisions that yield utility higher than the divider's proportionality guarantee; see Proposition \ref{proSymmetryBreaking}.

\subsection{Normal priors}\label{subComputationNormal}

\begin{algorithm}[t]
	\SetAlgoLined
	\DontPrintSemicolon
	\KwIn{Divider values $g^D_1, g^D_2, \dots, g^D_n$ (not all zero), prior means $\mu_1, \mu_2, \dots, \mu_n$, corresponding standard deviations $\sigma_1, \sigma_2, \dots, \sigma_n$, and an additive error bound $\gamma$ for the divider's optimal utility.}
	\KwOut{An approximately optimal division $\seq{p}{n}$.}
	$\delta \gets \frac{\gamma}{\sum_{i = 1}^n\abs{g^D_i}}$\;
	$P \gets \frac12$\;
	$u \gets -\infty$\;
	\While{$P > 0$}{
		$u_P, \seq{q}{i} \gets$ optimal solution to the following program $\mathcal{C}_P$, which has variables $\seq{q}{i}$:
		$$\lpmax{u_P = \sum\limits_{i = 1}^n\frac{g^D_i}{2} \left(P\left(1 - q_i\right) + (1 - P)\left(1 + q_i\right)\right)}{\\$-1 \leq q_i \leq 1$ & for all $1 \leq i \leq n$,\\$\sum\limits_{i = 1}^{n} g^D_i q_i \geq 0$,\\$\sum\limits_{i = 1}^{n} \mu_i q_i \leq \Phi\inv(P)\sqrt{\sum\limits_{i = 1}^{n} \sigma_i^2 q_i^2}$}$$
		\If{$u_P > u$}{
			$u \gets u_P$\;
			\For{$i \gets 1, 2, \dots, n$}{
				$p_i \gets \frac{q_i + 1}{2}$\;
			}
		}
		$P \gets P - \delta$\;
	}
	\Return{$(\seq{p}{n})$}\;
	\caption{Computes an approximately optimal division given the divider's values and independent normal priors for the chooser's values.}
	\label{algNormalOpt}
\end{algorithm}

We now turn to prove the first main result of this paper: an efficient algorithm to compute a near-optimal division given the divider's values for each good and normal priors for the chooser values. The procedure is presented formally as Algorithm \ref{algNormalOpt}; first we give an informal explanation. With a bit of manipulation, we can rewrite the divider's optimization problem as maximizing
$$\ee[u^D] = \sum\limits_{i = 1}^n\frac{g^D_i}{2} \left(P\left(1 - q_i\right) + (1 - P)\left(1 + q_i\right)\right)$$
over the variables $\seq{q}{n} \in [-1, 1]$. This is not a linear program. However, the only nonlinearities arise from the $P$ term, which is itself a function of the $q_i$. The key idea is to try to guess the optimal $P$ by trying several different values, uniformly spread out between 0 and $\frac12$. For each guessed value of $P$, we add a constraint that the chooser picks pile 1 with probability at most $P$. For normal priors, this turns the linear program into a quadratic program, yet it is still convex, so can be readily solved in polynomial time. The only catch is that we lose exact optimality, picking up an error term from potentially missing the exact optimal value of $P$. Fortunately, we can use the structure of the objective function to bound this error in a way that gives a very strong guarantee of approximate optimality.

In what follows, $\Phi$ denotes the standard normal cumulative distribution function.

\begin{lemma}\label{lemNormalAlgoPolytime}
	Algorithm \ref{algNormalOpt} runs in polynomial time in the values of $n$ and $\frac{\sum_{i = 1}^n\abs{g^D_i}}{\gamma}$.
\end{lemma}

\begin{proof}
	Observe that there are at most
	$\frac{1}{2\delta} = \frac{\sum_{i = 1}^n\abs{g^D_i}}{2\gamma}$
	iterations of the main loop. Thus, all that remains to show is that each iteration takes polynomial time in $n$. This follows from the observation that each $\mathcal{C}_P$ is a convex program. To see this, note that the objective function is clearly linear in the $q_i$ variables, and all constraints except for the final one are linear as well. The final constraint is not linear, but we claim that the set $S \subseteq \rr^n$ of points satisfying the constraint is convex. Suppose $q := (\seq{q}{n}) \in S$, and let $\widetilde{q} := (\sigma_1 q_1, \sigma_2 q_2, \dots \sigma_n q_n)$. Then, for any positive real number $c$, the scaled vector $cq = (\frac{q_1}{c}, \frac{q_2}{c}, \dots, \frac{q_n}{c})$ lies in $S$ as well, since
	\begin{align*}
		\sum\limits_{i = 1}^{n} \mu_i (c q_i) = c \cdot \sum\limits_{i = 1}^{n} \mu_i q_i
		&\leq c \cdot \Phi\inv(P)\sqrt{\sum\limits_{i = 1}^{n} \sigma_i^2 q_i^2} \snc{q \in S}\\
		&= \Phi\inv(P) c \abs{\abs{\widetilde{q}}}_2 \stext{where $\abs{\abs{\cdot}}_2$ denotes the Euclidean norm}\\
		&= \Phi\inv(P) \abs{\abs{c \widetilde{q}}}_2\\
		&= \Phi\inv(P)\sqrt{\sum\limits_{i = 1}^{n} \sigma_i^2 (cq_i)^2}.
	\end{align*}
	Therefore, $S$ is a cone centered at the origin, which is a convex set.
\end{proof}

We remark that, beyond being efficient in theory, this algorithm proves to be fast in practice. We implemented this algorithm using Gurobi \cite{Gurobi} to solve $\mathcal{C}_P$ as a convex quadratic program and used it to verify many of the examples in this paper.

\begin{lemma}\label{lemNormalAlgoCorrect}
	Algorithm \ref{algNormalOpt} finds a division yielding divider utility within an additive $\gamma$ of the optimal divider utility.
\end{lemma}

The proof is long and technical, so it is deferred to Appendix \ref{appProofNormalAlgoCorrect}.

With a bit more work, we can translate this additive approximation guarantee into a multiplicative one. Formally, we can show that Algorithm \ref{algNormalOpt} is a \emph{fully polynomial-time approximation scheme (FPTAS)} for maximizing divider utility, which means that, on instances where the optimal value is $u^* > 0$, for any $\varepsilon > 0$, it can find a solution with objective value at least $(1 - \varepsilon) \cdot u^*$ in time polynomial in both $n$ and $\frac{1}{\varepsilon}$.

\begin{theorem}\label{thmNormalAlgoFPTAS}
	When all $g^D_i \geq 0$, running Algorithm \ref{algNormalOpt} with $\gamma := \frac{\varepsilon}{2} \cdot \sum_{i = 1}^{n} g^D_i$ is a fully polynomial-time approximation scheme with approximation parameter $\varepsilon$.
\end{theorem}

\begin{proof}
	Given this choice of $\gamma$, it follows from Lemma \ref{lemNormalAlgoPolytime} and the fact that $\abs{g^D_i} = g^D_i$ that the algorithm runs in polynomial time in $n$ and $\frac{1}{\varepsilon}$. Let $u$ denote the utility of the solution returned by the algorithm, and let $u^*$ denote the optimal utility. By Lemma \ref{lemBaselineUtilities},
	$2u^* \geq \sum_{i = 1}^{n} g^D_i.$
	Combining this inequality with Lemma \ref{lemNormalAlgoCorrect}, we have
	$$u \geq u^* - \gamma = u^* - \frac{\varepsilon}{2} \cdot \sum_{i = 1}^{n} g^D_i \geq u^* - \frac{\varepsilon}{2} \cdot 2u^* = (1 - \varepsilon) \cdot u^*.$$
	as desired.
\end{proof}

\ipnc{.8}{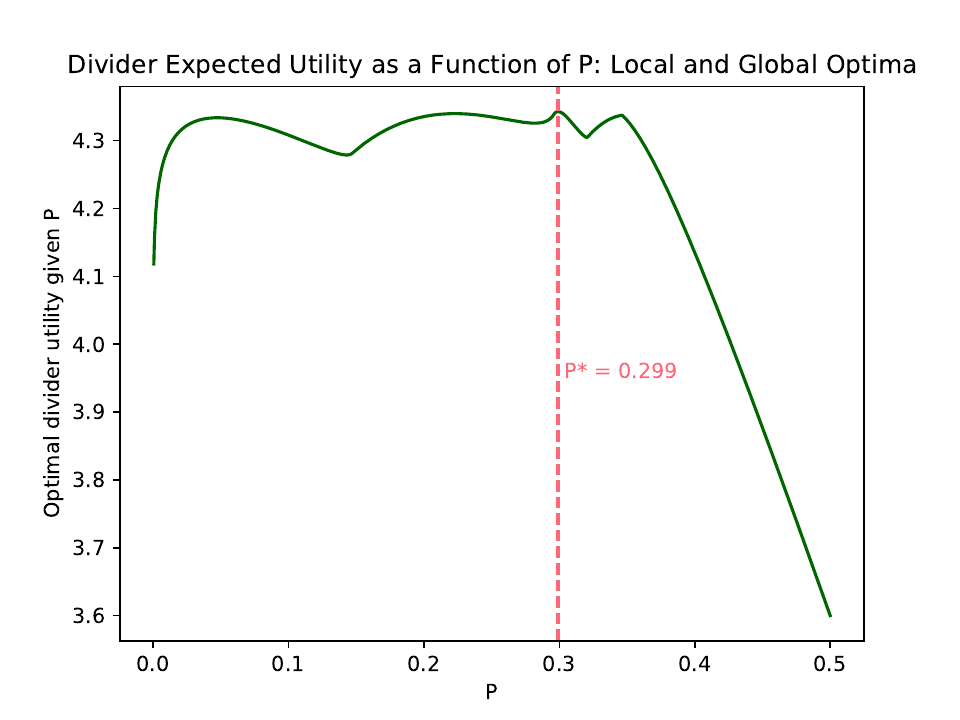}{\label{figManyPeaksExample}An instance with four goods where there are four locally-optimal divisions with a variety of different values of $P$. The globally optimal value of $P$ is indicated as $P^*$. The divider values are $g^D_1 = 3$, $g^D_2 = 2$, $g^D_3 = 1$, $g^D_4 = 1.2$, and corresponding chooser priors are $\overline{\mathcal{G}}^C_1 = \nrm{5, 1}$, $\overline{\mathcal{G}}^C_2 = \nrm{9.5, 1}$, $\overline{\mathcal{G}}^C_3 = \nrm{13.6, 96.04}$, $\overline{\mathcal{G}}^C_4 = \nrm{95, 28561}$.}

One might wonder why it is necessary to sequentially search for the optimal value of $P$. If the optimal divider utility given $P$ were a single-peaked function of $P$, then it would be possible to rapidly compute an optimal division through a ternary search over $P$. However, single-peakedness is not always satisfied, as Figure \ref{figManyPeaksExample} demonstrates. In fact, local optima can occur even in very simple scenarios:

\begin{proposition}\label{proLocalOptima}
	Even for $n = 3$ goods, there exist positive divider values and normal chooser priors with identical positive means such that the divider's interim expected utility, as a function of the division $p$, has a local maximum that is not a global maximum.
\end{proposition}

\begin{proof}
	This was discovered and verified via computational methods, so here we just explain the example at a high level. Suppose $\overline{\mathcal{G}}^C_1 = \nrm{100, 1}$, $\overline{\mathcal{G}}^C_2 = \nrm{100, 1}$, $\overline{\mathcal{G}}^C_3 = \nrm{100, 4225}$, and $g^D_1 = 11, g^D_2 = 9$, $g^D_3 = 1$. There are two locally-optimal strategies. Approximately, they are:
	\begin{enumerate}[label={(\roman*)}]
		\item\label{itmStrategyLowVariance} Divide the high-variance good 3 evenly between the two piles, so that it has no influence on the probability that the chooser picks pile 1. Then execute the optimal perfect-information strategy, putting most of good 1 in pile 1 and all of good 2 in pile 2. The risk that the chooser picks pile 1 is very low, at $P \approx 0.015$.
		\item\label{itmStrategyHighVariance} Put good 3 entirely into pile 2, so that it is possible to extract a substantial amount of both goods 1 and 2 in pile 1. The risk that the chooser picks pile 1 is moderate, at $P \approx 0.21$.
	\end{enumerate}
	As we verified using Algorithm \ref{algNormalOpt}, strategy \ref{itmStrategyLowVariance} yields utility of approximately 11 (as one would expect), whereas, despite the risk in relying on the high-variance good, strategy \ref{itmStrategyHighVariance} yields utility of approximately 12. However, intermediate strategies yield utilities less than either one of these extremes.
\end{proof}

\subsection{Symmetric log-concave priors}\label{subFPRAS}

It turns out that the technique from Algorithm \ref{algNormalOpt} applies to a much wider class of distributions. Specifically, we consider distributions $\mathcal{D}$ whose density functions have two properties:
symmetry about their mean and log-concavity. While we are not able to get a clean quadratic program, as we obtain in the case of normal distributions, we can still show convexity. As an application, we give a randomized approximation algorithm that handles the case where each $\overline{\mathcal{G}}^C_i$ is a uniform distribution over an interval.

Many well-studied families of continuous distributions are symmetric and log-concave, including normal distributions and uniform distributions over intervals. A product measure
$$\mathcal{D} = \mathcal{D}_1 \times \mathcal{D}_2 \times \dots \times \mathcal{D}_n$$
is symmetric and log-concave in an $n$-dimensional sense whenever each $\mathcal{D}_i$ is. Our interest in these distributions stems from the following theorem.

\begin{theorem}[Meyer and Reisner \cite{MeyerReisner2}]\label{thmMR2}
	Let $\mathcal{D}$ be a symmetric, log-concave measure over $\rr^n$ and let $0 \leq P \leq \frac12$. There exists a set $K_P \subseteq \rr^n$ such that, for any hyperplane $H$ separating $\rr^n$ into halfspaces $H^-$ and $H^+$,
	\begin{enumerate}[label={(\roman*)}]
		\item\label{itmMR2Iff} $\Pr_{x \sim \mathcal{D}}[x \in H^+] \geq P$ if and only if $H^+ \cap K_P$ is nonempty, and
		\item\label{itmMR2Centroid} $\Pr_{x \sim \mathcal{D}}[x \in H^+] = P$ if and only if $H^+ \cap K_P$ contains a single point, which is always the centroid of $\mathcal{D}$ restricted to $H$.
	\end{enumerate}
\end{theorem}

The sets $K_P$ are called ``floating bodies'' of $\mathcal{D}$. When $\mathcal{D}$ is a uniform distribution over a compact set $X \subseteq \rr^n$, a geometric interpretation lends insight. We consider all possible ways to slice off a $P$-fraction of the volume of $X$. After all such slices are made simultaneously, $K_P$ is the set that remains. Then the backward direction of statement \ref{itmMR2Iff} is immediate, following from the observation that every supporting hyperplane of $K_P$ slices off at least a $P$-fraction of the volume of $X$. The nontrivial forward direction is that there are no ``redundant'' hyperplanes: every plane slicing off at least a $P$-fraction of the volume of $X$ is a supporting hyperplane of $K_P$. This part requires the distribution $\mathcal{D}$ to be symmetric and log-concave. It implies that each hyperplane intersects $K_P$ at a single point, which must be the center of mass of $X \cap H$ by a straightforward calculus exercise.

The Meyer-Reisner Theorem has received surprisingly little attention in the literature on convex optimization. However, it is intimately tied to convexity, and plays a crucial role in the proofs of our next two theorems.

\begin{theorem}\label{thmSymmetricConvexity}
	If the chooser prior $\overline{\mathcal{G}}^C$ is symmetric and log-concave, then the set $S_P$ of divisions for which the probability the chooser prefers Pile 1 is at most $P$ is a convex set.
\end{theorem}

\begin{proof}
	Let $q^1, q^2 \in [-1, 1]^n$ be two divisions for which the probability the chooser picks Pile 1 is at most $P$. Suppose, toward a contradiction, that for some $0 \leq \alpha \leq 1$, the probability the chooser picks Pile 1 under $q^3 := \alpha q^1 + (1 - \alpha) q^2$ is $P' > P$. In other words, for $i \in \{1, 2\}$,
	\begin{equation}\label{equQ1Q2Ok}
		\Pr_{g^C \sim \overline{\mathcal{G}}^C}[q^i \cdot g^C \geq 0] \leq P,
	\end{equation}
	but
	\begin{equation}\label{equQ3NotOk}
		\Pr_{g^C \sim \overline{\mathcal{G}}^C}[q^3 \cdot g^C \geq 0] = P' > P.
	\end{equation}
	For each $i \in \{1, 2, 3\}$, consider the $(n - 1)$-dimensional plane
	\begin{align*}
		H_i := \{g^C \in \rr^n \suchthat q^i \cdot g^C = 0\},
	\end{align*}
	with positive side defined as
	\begin{align*}
		H_i^+ := \{g^C \in \rr^n \suchthat q^i \cdot g^C \geq 0\}.
	\end{align*}
	Then Equation (\ref{equQ3NotOk}) says that
	$$\Pr_{g^C \sim \overline{\mathcal{G}}^C}[g^C \in H_3^+] = P'.$$
	Applying Theorem \ref{thmMR2} part \ref{itmMR2Iff} (forward direction), we deduce that there is some specific point $g^C \in K_{P'} \cap H_3^+$. Since
	$$\alpha (q^1 \cdot g^C) + (1 - \alpha) (q^2 \cdot g^C) = q^3 \cdot g^C \geq 0,$$
	it follows that, for some $i \in \{1, 2\}$, we have $q^i \cdot g^C \geq 0$. Applying Theorem \ref{thmMR2} part \ref{itmMR2Iff} (backward direction) to $H_i$, it follows that
	$$\Pr_{g^C \sim \overline{\mathcal{G}}^C}[q^i \cdot g^C \geq 0] = \Pr_{g^C \sim \overline{\mathcal{G}}^C}[g^C \in H_i^+] \geq P',$$
	contradicting Inequality (\ref{equQ1Q2Ok}).
\end{proof}

Already, this suggests that the technique from Algorithm \ref{algNormalOpt} can be extended beyond normal distributions. When each $\overline{\mathcal{G}}^C_i$ is Gaussian, the constraint that $q \in S_P$ is a convex quadratic constraint. If we only require $\overline{\mathcal{G}}^C_i$ to be symmetric and log-concave, the constraint is no longer necessarily quadratic, but it is at least convex.

However, convexity alone does not immediately imply we can optimize over $S_P$ in polynomial time. For that we must build a \emph{separation oracle}, which is an algorithm that can efficiently decide whether any given point $q$ is in $S_P$, and if it is not, output a hyperplane separating $q$ from $S_P$, i.e., a linear constraint that every point in $S_P$ satisfies but $q$ does not. It turns out that neither of these tasks is computationally tractable in general. In the case where each $\overline{\mathcal{G}}^C_i$ is a uniform distribution over an interval, determining whether a given point is in $S_P$ is equivalent to computing the volume of an $(n - 1)$-dimensional plane cuts out of an $n$-dimensional cube. That well-studied problem is thought to be intractable.\footnote{In general, computing volumes of polyhedra is \#P-hard \cite{VolumeHardness2}. The best known formula for this particular problem involves enumerating the exponentially-many vertices \cite{HyperplaneSections}.} However, our next result shows that we can effectively solve this problem by the means of sophisticated random sampling techniques, using Theorem \ref{thmMR2} part \ref{itmMR2Centroid} to get an (approximate) separating hyperplane. Since the algorithm uses randomness, it is not an FPTAS but rather an FPRAS (fully polynomial-time \emph{randomized} approximation scheme). An FPRAS is the same as an FPTAS except that it is only required to be correct with some constant probability greater than $\frac12$.

\begin{theorem}\label{thmUniformOpt}
	An FPRAS exists to compute an approximately optimal division given rational divider values $g^D_1, g^D_2, \dots, g^D_n$ and chooser priors $\overline{\mathcal{G}}^C_i = \textup{Unif}[a_i, b_i]$ encoded by rational numbers $a_1, a_2, \dots, a_n$ and $b_1, b_2, \dots, b_n$.
\end{theorem}

The proof is deferred to Appendix \ref{appFPRAS}.

\subsection{Discrete priors}\label{subComputationDiscrete}

Using similar ideas we can also efficiently solve the case where the chooser's prior is a discrete distribution when the number of chooser types is small.

\begin{proposition}\label{proDiscreteAlgoCorrect}
	There is a function $f$ such that, given a discrete joint distribution $\overline{\mathcal{G}}^C$ over the chooser's values (even allowing for correlations across goods), with $n$ goods and $\abs{\overline{\mathcal{G}}^C} = \ell$, it is possible to compute an optimal division in time $f(\ell) \cdot \textup{poly(n)}$.
\end{proposition}

Algorithm \ref{algDiscreteOpt}, presented in Appendix \ref{appProofDiscreteAlgoCorrect}, solves this problem by optimizing a similar ensemble of linear programs and taking the best solution. Unlike our other algorithms, this returns an exact (rather than approximate) solution and works even when there are correlations among the chooser's values for the various goods.

\section{Diversification: Why and how}\label{secDiversification}

To maximize his expected utility, the divider must balance two objectives when he allocates goods to the piles: maximizing the returns from the more-desirable pile 1, and reducing the risk that the chooser picks pile 1. The divider trades off between these objectives by transferring just the right amount of value into pile 2. One might naturally expect that the divider's strategic approach from the case of known chooser preferences would still be optimal. Namely, the divider could start with all goods in pile 1 and transfer goods into pile 2 in order of ascending critical ratios, thus creeping along the risk-return frontier to find the optimal utility. In the end, at most one good would need to be divided.

However, another strategic factor---diversification---can reduce the risk to the divider that the chooser will pick pile 1, and thus enable the divider to do better. In the investment context, investors know they can push the entire risk-return frontier outwards by investing in many assets. This analogy applies quite well to our setting: given that the chooser is more likely to pick pile 2 (which is always the case by Lemma \ref{lemStructureGeneral}), when the expected difference in the chooser's values for the two piles is fixed, she is more likely to pick pile 2 if the variability of the difference is lower. Thus, the optimizing divider should not merely transfer a sufficient amount of value into pile 2. He should also reduce variance in this value by dividing multiple goods between the two piles, thereby diversifying the piles to reduce risk. In this section, we analyze how this incentive affects the divider's optimization problem.

If goods are indivisible, diversification can still be achieved by using lotteries, whereby the divider does not merely divide the goods into two piles, but commits to random allocations of certain goods after the chooser has made her choice. Consider goods 1 and 2 (of many others), for which the chooser's values are equally likely to be 0 or 1. If good 1 is put in pile 1 and good 2 in pile 2, then there is a 25\% chance that the chooser's value of those two assets will be greater in pile 1 than in pile 2. By contrast, if each good is put in pile 1 with a 50\% chance, and pile 2 with a 50\% chance, then the chooser---before the lotteries have been conducted---will always value the probabilistic assets in the piles equally. We emphasize here that the “diversification” imperative under examination is not aimed at reducing the risk in the lotteries for each good, but instead at reducing the risk that the chooser picks pile 1 before the lotteries are resolved. If the divider is risk-averse, then the incentive to diversify over risk in the chooser's action can be at odds with an inherent aversion to using lotteries. This is a subtle issue we address briefly in Section \ref{subDiversificationRiskAversion}, when we discuss risk-aversion given both divisible goods and indivisible goods that can be divided via lotteries.

\subsection{Which goods get divided?}\label{subDiversificationGeneral}

We know that the divider can find the optimal division of goods in the normal and discrete cases using Algorithm \ref{algNormalOpt} and Algorithm \ref{algDiscreteOpt} respectively. Here we provide a qualitative explanation of how goods are optimally allocated between the two piles.

\ipnc{.8}{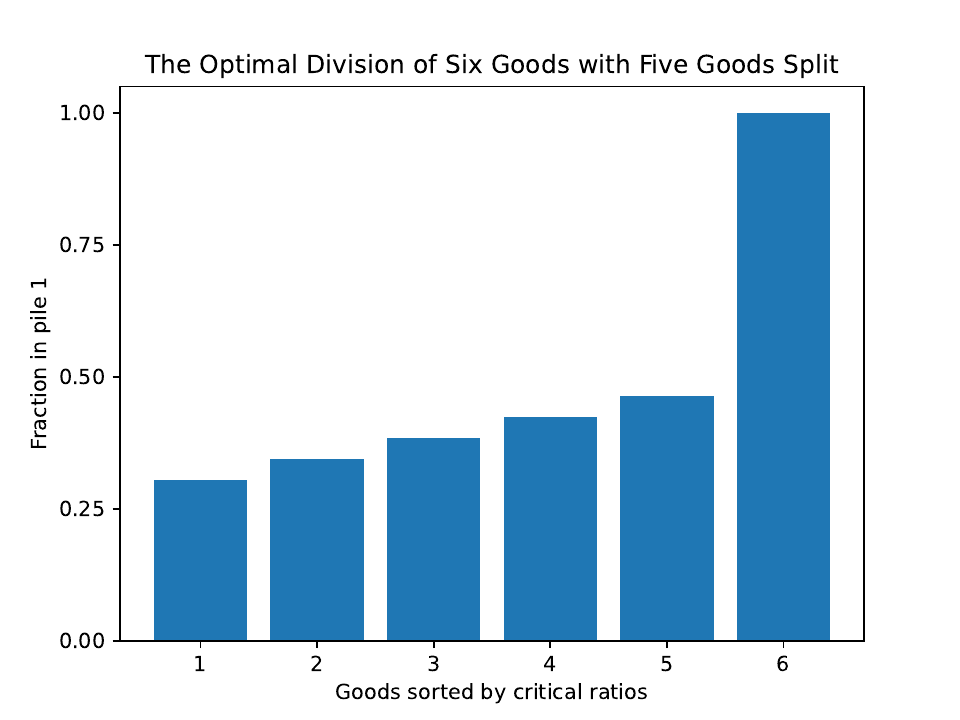}{\label{figManyGoodsDivided}An example with six goods where it is optimal for the divider to split five of the goods between the two piles. Here $g^D_1 = 9.8$, $g^D_2 = 9.9$, $g^D_3 = 10$, $g^D_4 = 10.1$, $g^D_5 = 10.2$, $g^D_6 = 15$, and the chooser's prior for each good is $\overline{\mathcal{G}}^C_i = \nrm{10, 1}$. The optimal value of $P$ is 0.078.}

We begin by observing that it may be optimal for the divider to split \emph{all but one good}, as Figure \ref{figManyGoodsDivided} illustrates. The most valuable good---worth about twice as much as any of the others---is placed entirely in pile 1, while the other goods are all placed mostly in pile 2 but split between the two piles in order to reduce variance, pushing $P$ extremely close to zero. This optimal division was computed using Algorithm \ref{algNormalOpt}.\footnote{Technically speaking, the division output by Algorithm \ref{algNormalOpt} is not guaranteed to be ``close'' to the globally optimal division, even though the objective values must be similar. However, for the divisions in Figures \ref{figManyGoodsDivided}, \ref{figMonotonicityViolation}, and \ref{figDiversificationAndRiskAversion}, we did verify that the optimal division, $p^*$, must be close to the computed division shown in the figure, $p$, in the sense that $\abs{\abs{p^* - p}}_\infty \leq 0.05$. (For Figure \ref{figDiversificationAndRiskAversion}, we only obtained 0.1 instead of 0.05.) This is a close enough approximation to conclude that the properties we are claiming hold in these two examples. We computed this by re-running Algorithm \ref{algNormalOpt} with an additional constraint that each $p_i$ be bounded away from the value in the original computed solution by 0.05. We did this separately for each good $i$, and each direction of the deviation (bounded away from above/below). For a small enough value of the error parameter $\gamma$, we can conclude that, in each of these $2n$ constrained optimizations, the optimal objective value decreases by more than $\gamma$. Thus, the globally optimal solution $p^*$ cannot respect any of these additional constraints, implying that $\abs{\abs{p^* - p}}_\infty \leq 0.05$.}

This result starkly contrasts with the case of complete information. As we discussed in Section \ref{subLitReview}, a divider with perfect information \emph{never} needs to split more than one good between the piles. Figure \ref{figDiversificationVersusCertainty} interpolates between these two cases by decreasing the variance from one down to zero. One by one, the high value goods are placed entirely in pile 1 and the low value goods are placed entirely in pile 2. Eventually, as the variance approaches zero, we arrive at the certainty case, where there is no longer any diversification.

\ipnc{.8}{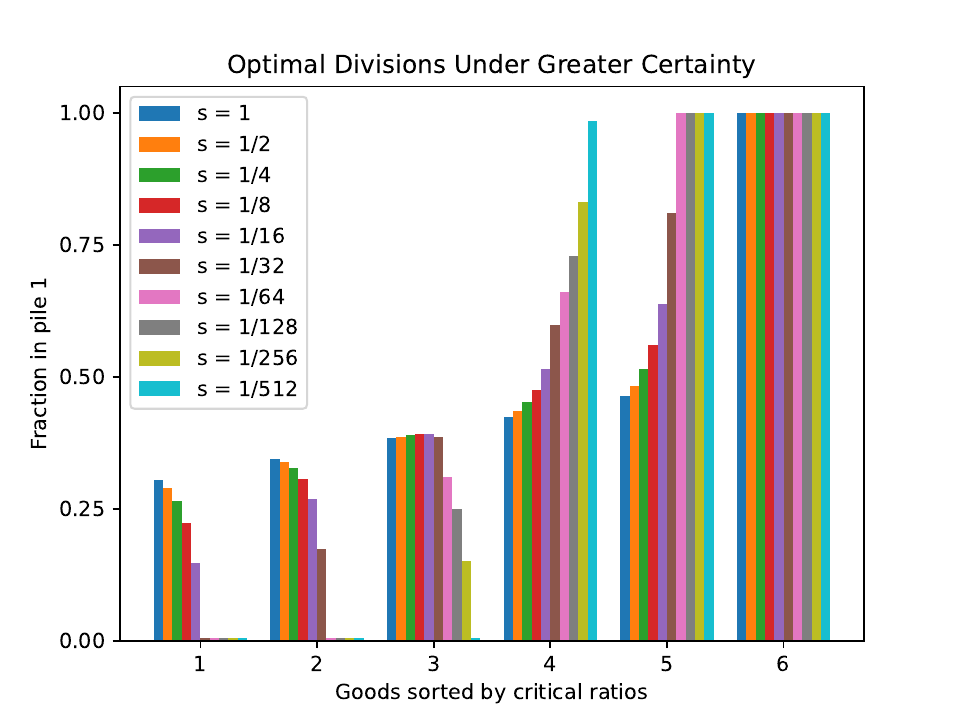}{\label{figDiversificationVersusCertainty} Optimal divisions as a function of the variance $s$ of the chooser's value for each good. This is the same setup as in Figure \ref{figManyGoodsDivided}, but instead of chooser values being drawn from $\nrm{10, 1}$, they are drawn from $\nrm{10, s}$, for $s$ starting at one and approaching zero. Thus the first (dark blue) bars are the same division as before.}

Although the first five goods are diversified in this example, the ones the divider values more tilt more toward pile 1 than do the ones the divider values less. This suggests an intuitive rule by which the divider could be guided in constructing his optimal division. Recall that, in the absence of uncertainty, the fraction $p_i$ of good $i$ placed in pile 1 is monotonically increasing in the critical ratio of good $i$.\footnote{Indeed, that fraction is zero or one, except for a single good intended to keep the piles even in value to the chooser.} Does this monotonicity result survive in the presence of uncertainty?

While it seems intuitive that prioritization by critical ratios would carry over to the uncertainty case, it will not if the priors have different variances. Even if a good has a very large or very small critical ratio, its variance may be so large to make it too risky to place it mostly in one pile. Thus, when variances differ among goods, monotone divisions may be suboptimal. Surprisingly, we observe that monotonicity may be lost even when goods have normal priors with \emph{identical} variances.

\ipnc{.8}{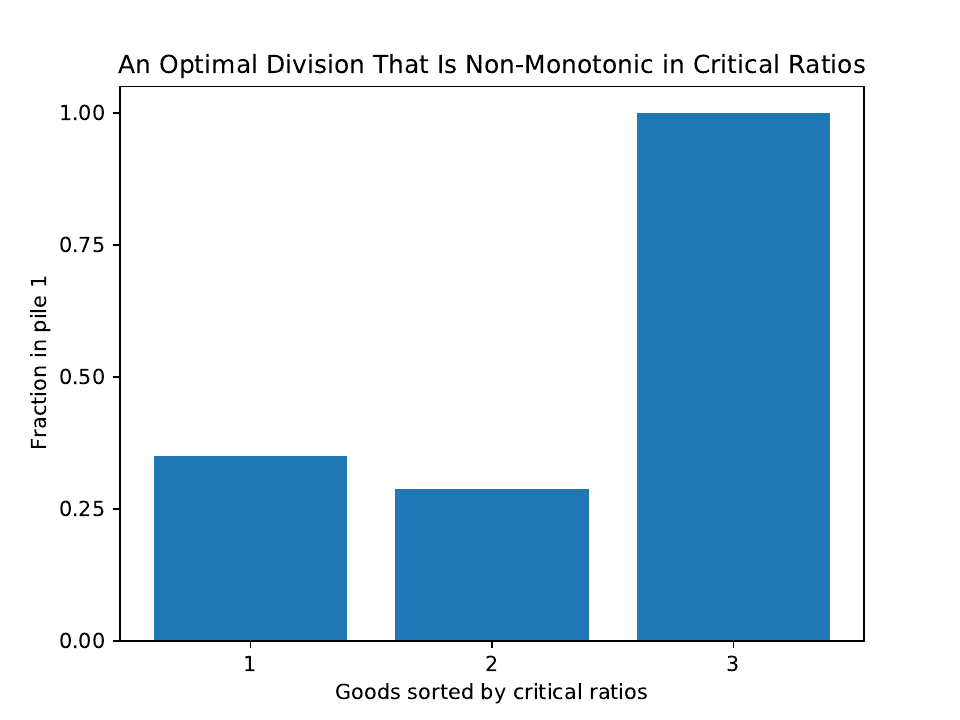}{\label{figMonotonicityViolation}An instance with three goods where, despite all prior variances being the same, the $p_i$ are not monotone in the critical ratios. Here $g^D_1 = 1$, $g^D_2 = 2$, $g^D_3 = 3$, and corresponding chooser priors are $\overline{\mathcal{G}}^C_1 = \nrm{100, 25}$, $\overline{\mathcal{G}}^C_2 = \nrm{198, 25}$, $\overline{\mathcal{G}}^C_3 = \nrm{100, 25}$. The optimal value of $P$ is 0.005.}

Figure \ref{figMonotonicityViolation} shows the optimal division in a simple example with three normally-distributed goods, again computed using Algorithm \ref{algNormalOpt}. Even though good 2 has a higher critical ratio and the same variance as good 1, the optimal division sets $p_2 < p_1$. The primary reason is that the value of good 2 to the chooser, in absolute terms (rather than relative to the divider's value), is so large that, in order to ensure that the value of $P$ stays low, raising $p_2$ would require lowering $p_1$ by such a large amount that the divider would be worse off. Note that we do require a high variance of 25 for each good to see this effect. As we decrease the variance to zero, we see convergence to the case of known preferences, with the division becoming monotone in the critical ratios starting around a variance of 5.

This example is surprising given the small amount of variance relative to the mean of each good. While in the polar case of zero variance, monotonicity in the critical ratios is optimal, a sliver of uncertainty breaks this result.

A further departure from the case of certain preferences is that the divider can achieve a higher-than-proportional utility even when all critical ratios are the same---in fact, even when the chooser's prior distributions are identical and the divider values every good identically. Clearly, such a division is not feasible with normal priors by Proposition \ref{proStructureNormal}. With two-point (or multi-point) discrete priors, despite such identical values across goods, it is feasible for the divider to beat his proportionality guarantee.

\begin{proposition}\label{proSymmetryBreaking}
	There exists a number of goods $n$ and a discrete distribution $\mathcal{D}$ supported on two positive values such that, even if, for each good $i \in [n]$, $g^D_i = 1$ and $\overline{\mathcal{G}}^C_i = \mathcal{D}$, there exists a division of goods yielding utility higher than the divider's proportionality guarantee.
\end{proposition}

\begin{proof}
	Let $n = 5$ and let $\mathcal{D}$ be the following distribution: value 0.01 with probability 0.6 and value 1 with probability 0.4. Consider the division
	$$p = (1, 0.4, 0.4, 0.4, 0.4).$$
	Observe that if the chooser values the first good at 0.01 and at least one of the other goods at 1, then she prefers pile 2. This happens with probability
	$$0.6 \cdot (1 - 0.6^4) = 0.553344 > \frac12.$$
	Since the divider values pile 1 more than pile 2, he achieves a higher utility than his proportionality guarantee; specifically, his expected utility is $2.510068 > 2.5$.
\end{proof}

Finally, we note that there is a limit to the need to diversify through lotteries. It is never strictly optimal to split \emph{all} goods between the two piles. The proof is straightforward, and is included in Appendix \ref{appOneGoodNotDivided} for completeness.
\begin{lemma}\label{lemOneGoodNotDivided}
	If the divider can achieve a utility exceeding his proportionality guarantee, he will always leave one good entirely in one of the two piles.
\end{lemma}

We also remark that, even though all examples above place an undivided good optimally in pile 1, sometimes it is optimal to place a sole undivided good in pile 2.

\subsection{The effects of correlation between the players' values}\label{subDiversificationCorrelation}

Thus far we have only considered settings where the divider's prior over the chooser's values is specified arbitrarily and is independent of the divider's values. In this section we briefly study optimal divisions embedded in a larger, realistic context where there is a joint prior over both players' values. Such a situation would arise if there were a public value component to the player's valuations, as say for market value when jewelry pieces or real estate assets are being divided. The divider uses his own values to update his beliefs about the chooser's values.

Specifically, consider a scenario where both players' values follow a Gaussian marginal distribution $\mathcal{G}^D_i = \mathcal{G}^C_i = \nrm{10, 1}$ for each good $i$. However, the values for each good are correlated. For each good $i$, there is a common public value $g_i$ that is distributed according to $\nrm{10, t}$, for a parameter $t \in [0, 1]$. Additionally, each player has a private value $\varepsilon^D_i, \varepsilon^C_i$ distributed independently according to $\nrm{0, 1 - t}$. The values for good $i$ are $g^D_i := g_i + \varepsilon^D_i$ and $g^C_i := g_i + \varepsilon^C_i$.

\ipnc{.8}{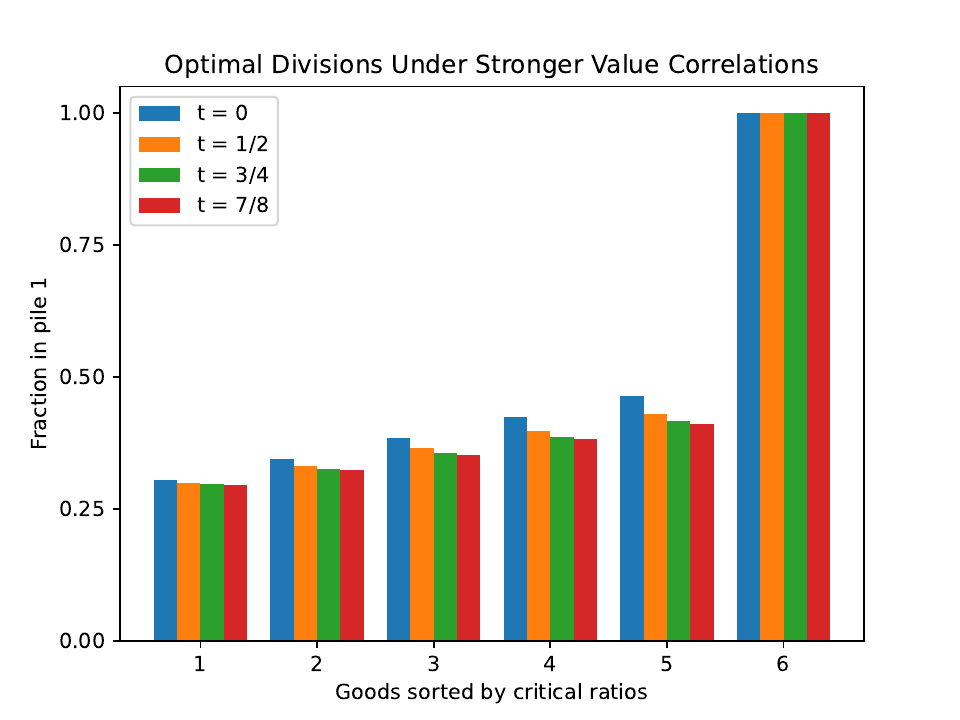}{\label{figDiversificationVersusCorrelation} Optimal divisions as a function of the correlation parameter $t$. As in Figure \ref{figManyGoodsDivided}, each value is is drawn from $\nrm{10, 1}$, and the divider's values happen to be $g^D = (9.8, 9.9, 10, 10.1, 10.2, 15)$. The difference is that now, for $t > 0$, the divider infers that the chooser has similar values to his own. The first (dark blue) bars corresponding to $t = 0$ are the same division as before. As $t$ approaches 1, the optimal division changes little from the final (red) bars at $t = 7/8$.}

Figure \ref{figDiversificationVersusCorrelation} shows the optimal divisions for increasing values of $t$ in the same setting as before, where the divider has one good at a very high value of 15 and the other goods' values are close to the mean of 10. As is evident, the divider's general strategy does not change in any significant qualitative way. The high-value good is still placed entirely in pile 1, and the divider still diversifies among the other goods. Perhaps surprisingly, as $t$ increases toward 1, the divider actually diversifies more, with the gaps between the fractions in pile 1 among the first 5 goods shrinking.

One possible explanation for this trend is as follows. As $t$ increases, the divider increasingly expects the chooser to have similar values, as most of the variation can be attributed to the public value. It thus becomes more important for the divider to divide ``evenly.'' Otherwise, if the divider concentrates too much value in pile 1, then because the chooser has similar values, the chooser will pick pile 1. But then, as the piles start to have similar values to the divider, the consequences of the chooser picking pile 1 become less dire. So the divider will optimally choose a division with a higher value of $P$. (Indeed, this is the case in this example. In going from $t = 0$ to $t = 1/2$, the value of $P$ doubles, then continues to approach $1/2$ as $t$ approaches 1.) In this regime, diversification becomes more important, as decreasing variability in the values of the two piles can have an even greater effect on improving the chances the chooser picks pile 2.

\subsection{The effects of risk aversion}\label{subDiversificationRiskAversion}

It is optimal for the divider to increase his expected value by taking a risk on the value he receives; how does his strategy adjust if he is averse to risk? Suppose the divider is an expected utility maximizer with utility $u^D = f(v^D)$, where $f$ is an increasing concave function and $v^D$ is the total value of all goods the divider receives.

Thus far, we have not distinguished between deterministic divisions of divisible goods versus randomized divisions of indivisible goods. For example, if $p_i = \frac12$, then it could mean that good $i$ is literally split between the piles into two equal pieces, or that it will be randomly allocated, in whole, to one pile or the other after the chooser picks a pile. In the two cases the two players' incentives are the same. However, if the divider is risk-averse, a lottery that is resolved after piles are selected will impose unwanted risk on him. In this section, we assume that the goods are divisible and note where results would be different if the goods were indivisible and allocated by lottery.

We identify two main effects of a divider's risk-aversion. First, it decreases the probability $P$ that the chooser picks pile 1, as the following theorem shows.

\begin{theorem}\label{thmRiskAversion}
	Assume all values of goods are always nonnegative. Fix divider values and let $f$ be a concave utility function. The probability that the chooser picks pile 1 under any optimal division by a risk-neutral divider is weakly greater than the probability the chooser picks pile 1 under any optimal division by a divider with utility function $f$.
\end{theorem}

\begin{proof}
	Let $p$ be the optimal division by a risk-neutral divider, and let $p'$ be the optimal division by a risk-averse divider with utility function $f$. Let $T$ be the total divider value of all goods. Let $v$, $P$, and $v'$, $P'$ denote the divider's value for pile 1 and the probability the chooser picks pile 1 according to $p$ and $p'$, respectively. Suppose toward a contradiction that $P < P'$. Then we must have $v < v'$, for otherwise $p$ would be a strictly better division than $p'$ for the risk-averse divider, as it would simultaneously yield a higher value in pile 1 and higher probability of the chooser picking pile 2. Also, since the risk-neutral divider prefers $q$ to $q'$, it must be that
	$$(1 - P')v' + P'(T - v') < (1 - P)v + P(T - v).$$
	
	\ipnc{1}{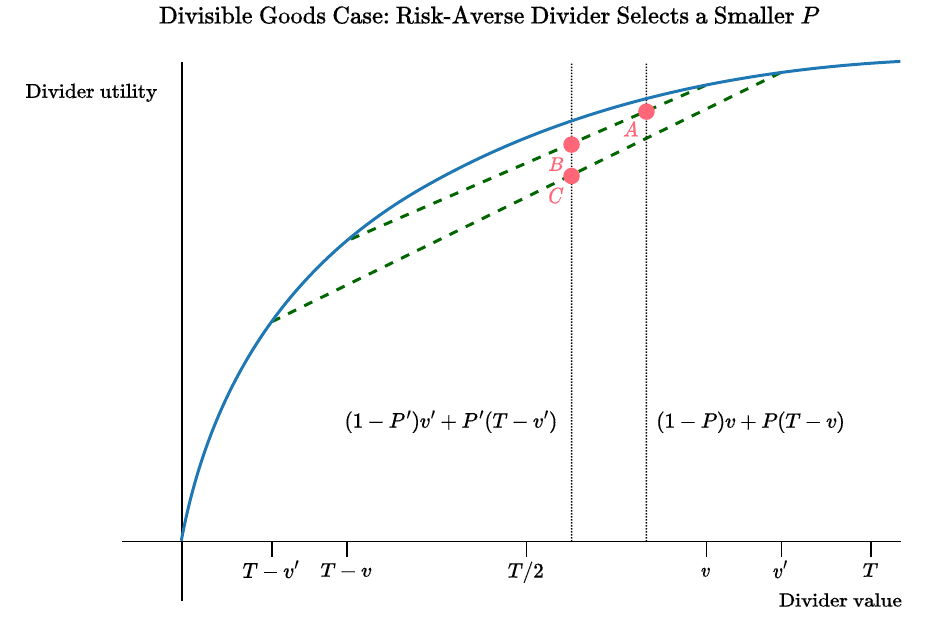}{\label{figRiskAversionPictorialProof}Illustration accompanying the proof of Theorem \ref{thmRiskAversion}.}
	
	Thus, values are ordered exactly as shown in Figure \ref{figRiskAversionPictorialProof}, where the $x$-coordinate of points $A$ and $C$ are the respective expected divider values of $p$ and $p'$. If the solid blue curve is the utility function $f$, then the $y$-coordinates of points $A$ and $C$ are the expected utilities of the risk-averse divider using $p$ and $p'$, respectively. It follows from monotonicity and convexity that $A$ must have a higher $y$-coordinate than $B$, which must have a higher $y$-coordinate than $C$. This contradicts our assumption that the risk-averse divider prefers $p'$.
\end{proof}

We may observe this effect empirically by assuming the utility function $f(x) = \sqrt{x}$. This turns the convex program $\mathcal{C}_P$ from Algorithm \ref{algNormalOpt}, which maximizes divider utility with respect to a fixed value of $P$, into a non-convex program. However, it can be written with only two non-convex constraints, so it is still practically feasible to solve it exactly. Figure \ref{figLPulldown} plots the optimal expected divider utility given any value of $P$ for a divider with utility function $f(x) = \sqrt{x}$. The divider's values and chooser's priors are the same as in Figure \ref{figManyPeaksExample}. Comparing the two figures, one can see that risk aversion reduces the optimal utility for larger values of $P$, making smaller values of $P$ more attractive.

\ipnc{.8}{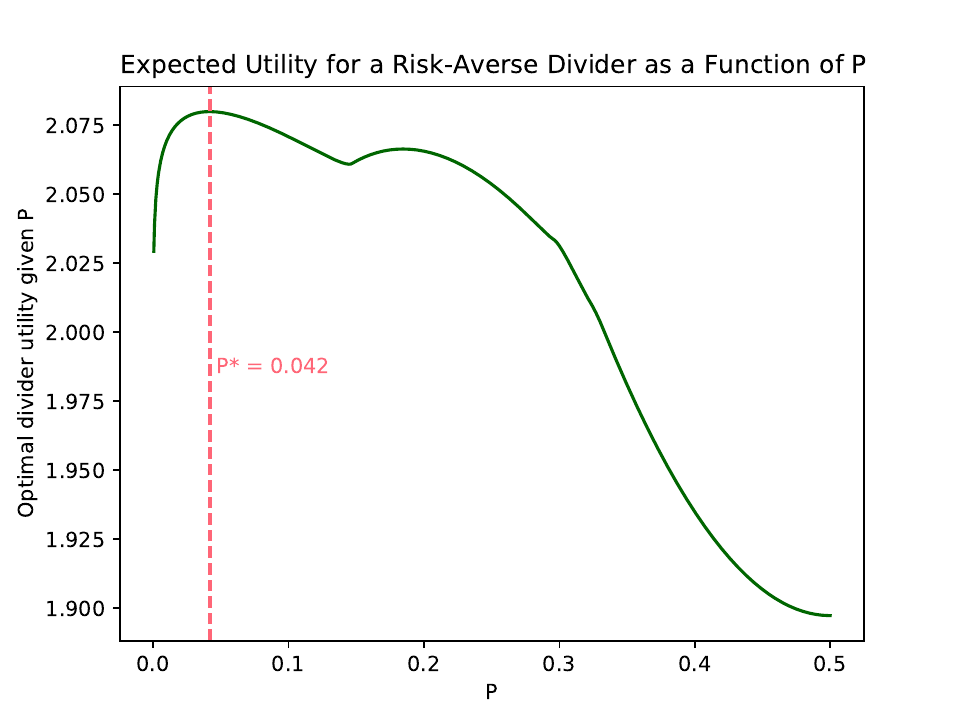}{\label{figLPulldown}The same plot of expected divider utility versus the probability that the chooser picks pile 1 as in Figure \ref{figManyPeaksExample}, but with a risk averse divider. This example uses the same divider values and chooser priors as in Figure \ref{figManyPeaksExample}. Again, the globally optimal value of $P$ is indicated as $P^*$.}

Theorem \ref{thmRiskAversion} does not hold when goods are indivisible and allocated randomly. Consider a scenario with only $n = 2$ goods. Good 1 deterministically has value 4 for both players. Good 2 is worth 16 to the divider and either 1 or 12 to the chooser, each with probability $\frac12$. A risk-neutral divider will put $\frac23$ of good 2 in pile 1 and everything else in pile 2, ensuring that the chooser always picks pile 2 (i.e., $P = 0$). If the divider has utility function $f(x) = \sqrt{x}$, then assuming ``$\frac23$ of good 2'' is a lottery over good 2, this strategy yields expected utility $\frac23 \sqrt{16} = \frac83$. However, by just putting the goods in separate piles and not using lotteries, the chooser will pick pile 2 with probability $P = \frac12$, yielding expected divider utility
$$\frac12 \sqrt{4} + \frac12 \sqrt{16} = 3 > \frac83.$$
We verified computationally that no division in which $P = 0$ yields expected divider utility greater than $\frac83$.

The second major effect of divider risk aversion is to increase the amount of diversification. Consider the setting with 40 goods with both divider and chooser values drawn i.i.d.\txt{} from $\nrm{1, 0.04}$. Figure \ref{figDiversificationAndRiskAversion} compares the optimal divisions by a risk-neutral divider and a risk-averse divider using the utility function $f(x) = \sqrt{x - 5}$ with values for the 10 goods drawn independently from $\nrm{1, 0.04}$. As one can see, risk-aversion leads to more goods being split between the two piles, and they are generally split more equally. Still, one good is always left undivided, as Lemma \ref{lemOneGoodNotDivided} holds with the same proof.

\ipnc{.7}{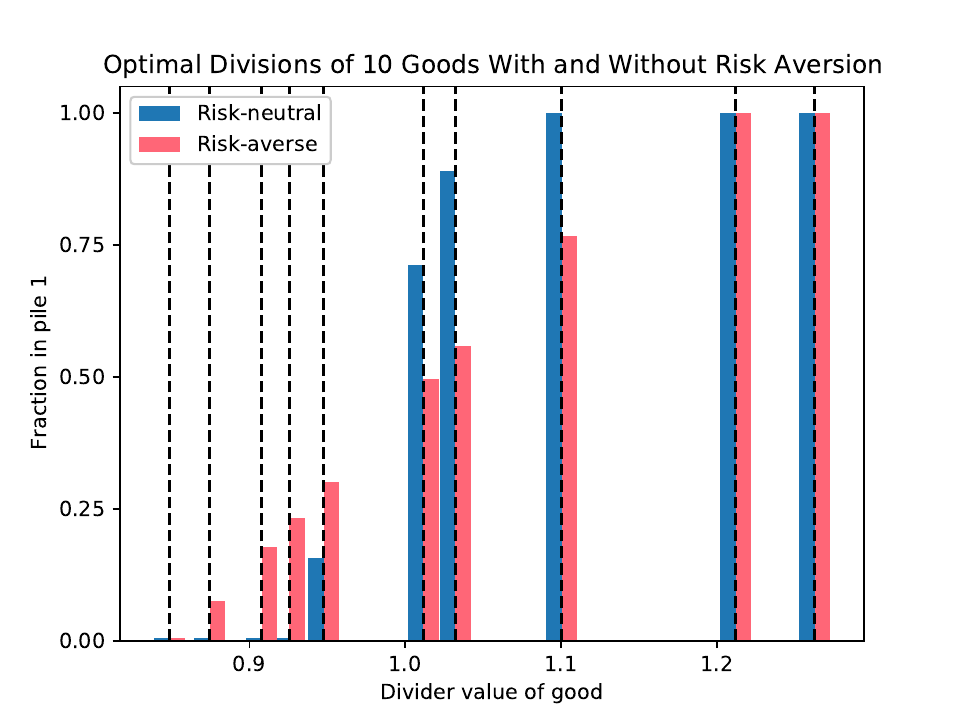}{\label{figDiversificationAndRiskAversion}Comparison of the divider's optimal divisions under risk-neutrality (left, dark blue bars) and risk-aversion (right, light red bars). The chooser's prior for each good $i$ is $\mathcal{G}^C_i := \nrm{1, 0.04}$, and the divider's values were also independently sampled from $\mathcal{G}^D_i := \nrm{1, 0.04}$. Each bar represents a value of $p_i$, and is horizontally located at the divider's value $g^D_i$ (indicated by the dashed black lines).}

\section{Welfare of the players}\label{secWelfare}

We now turn to analyze the expected welfare of the players. Knowing their expected welfare is important if our concern is some conception of fairness. It is also critical in enabling a player to decide when to try to play divider, and when chooser. The divide and choose game is explicitly asymmetric, and despite its minimal axiomatic guarantees (e.g., envy-freeness), one player might end up better off than the other merely due to this asymmetry. In this section we identify conditions under which one role is preferable to the other.

\subsection{Number of goods}\label{subWelfareDividerVersusChooser}

The cake-cutting literature suggests that the chooser is better off for the unknown-preferences case, as the divider is compelled to divide goods roughly evenly, while the chooser can get a more favorable outcome because she will typically value one pile more than the other. However, if the divider has strong knowledge of the chooser's preferences, then the divider can exploit this knowledge \cite{HPYoungBook, AlgorithmicFramework}.  Indeed, as Nicol\'o and Yu \cite{ArielRef} note in the context of cake-cutting, ``The divide and choose rule leads to a no-envy outcome but the rule itself is not envy free: the chooser envies the role of the divider.'' Thus, the best generalization one could hope to make is that the relative utilities of the two players depend on the amount of uncertainty faced by the divider. Hence, the greater one's own and one's counterpart's knowledge, the greater the benefit of playing the divider. Weak knowledge, on either side, favors being the chooser.

One natural example of this phenomenon is when values for all $n$ goods are drawn i.i.d.\txt{} from the same distribution. If $n$ is small, there is significant uncertainty in how the chooser will value the piles. Moreover, the divider cannot count on receiving what he places in pile 1. Consequently, when goods are few, the chooser has the advantage.

In contrast, when $n$ is large, the uncertainty in the value of a pile shrinks relative to its mean value. In this situation, the divider can cluster his high-value goods into pile 1, and expect to receive that pile with high probability. Thus, the divider is favored. These observations lead to the final major result in this paper: For plausible distributions of players' values, the chooser is favored when $n$ is small; the divider is favored when $n$ is large. In the latter case, this is true even when values are correlated (positively or negatively) between the players.

\begin{theorem}\label{thmCrossover}
	Let $\mathcal{D}$ be a probability distribution over $\rr_{> 0} \times \rr_{> 0}$. Suppose the values for each good $i$ are drawn from $\mathcal{D}$, independently across $i$. Then, as long as the critical ratio of each good is not always the same value with probability one, the following hold in any equilibrium.
	\begin{enumerate}[label={(\roman*)}]
		\item\label{itmCrossoverTheorem2Goods} If the two components of $\mathcal{D}$ are independent, then for $n = 2$ it is strictly better to be the chooser:
		$$\ee_{\substack{(g^D, g^C) \sim \mathcal{D}^n}}[u^D] < \ee_{\substack{(g^C, g^D) \sim \mathcal{D}^n}}[u^C]$$
		\item\label{itmCrossoverTheoremManyGoods} If,  for all $\varepsilon > 0$, for all sufficiently large $n$,
		\begin{equation}\label{equCrossoverSubexpTailAssumption}
			\Pr_{(g^1, g^2) \sim \mathcal{D}}[g^2 \leq 8n \exp(-2\varepsilon^2 n)] < \frac{\varepsilon}{n},
		\end{equation}
		then for all sufficiently large $n$, it is strictly better to be the divider:
		$$\ee_{\substack{(g^D, g^C) \sim \mathcal{D}^n}}[u^D] > \ee_{\substack{(g^C, g^D) \sim \mathcal{D}^n}}[u^C]$$
	\end{enumerate}
\end{theorem}

As an illustration of Theorem \ref{thmCrossover}, take $\mathcal{D}$ to be the uniform distribution on $[0, 1]^2$: every player's value for every good is drawn uniformly from the unit interval. This easily satisfies all the assumptions of Theorem \ref{thmCrossover}. Table \ref{tabU01Utilities} lists the ex-ante expected utilities for each player with two goods and with many goods. With two goods, the chooser is about 50\% better off; with many goods, the divider is 50\% better off.

Why does this happen? In the two-good case, the key observation is that the divider will always fractionally allocate his favorite good, mostly in pile 1, placing his less favorite good entirely in pile 2. Thus, when the chooser's favorite good is different from the divider's, she always does better, since she receives all of the good rather than part of it. When they prefer the same good, one can likewise see that the chooser is still better off via a simple case analysis over the relative fractions of the favorite good that each player would have kept in pile 1 had they been the divider.

For large $n$, relying on the law of large numbers, the divider could be confident that if he placed slightly more than 50\% of the goods in pile 2, the chooser would pick pile 2 with an arbitrarily high probability. For example, when $n = 100$, on average, the optimizing divider puts his 45 highest-valued goods in pile 1, and his 55 lowest-valued goods in pile 2. The chooser picks pile 1 with probability $P = 0.04$. In the limit as $n \to \infty$, $P$ tends to zero, so the divider's ex-ante utility is the sum of his top-half most-valued goods. On the other hand, for very large $n$, by the law of large numbers, the chooser can expect only her proportionality guarantee utility from each pile, i.e., the sum of a \emph{random} subset of half of her values. This is essentially the whole argument of the proof, though establishing the trends in full generality, even when values are correlated, requires carefully defining and analyzing the optimal divisions.

\begin{center}\begin{table}[H]\centering\begin{tabular}{r|c|c}
			Equilibrium utility per $U[0, 1]$ good & $n = 2$ & $n \to \infty$ \\\hline
			Proportionality guarantee & 0.25 & 0.25\\
			Divider & $\frac{19}{72} = 0.264$ & 0.375 \\
			Chooser & $\frac{1031}{2160} + \frac{\ln(3/4)}{3} = 0.381$ & 0.25
		\end{tabular}\caption{\label{tabU01Utilities}Expected utilities of the two players, normalized by the number of goods, when the value each player has for each good is drawn i.i.d.\txt{} from the uniform distribution on $[0, 1]$. The $n = 2$ column was computed using Mathematica. The $n \to \infty$ column was computed using the proof of Theorem \ref{thmCrossover} \ref{itmCrossoverTheoremManyGoods}.}\end{table}\end{center}

In Appendix \ref{appCrossoverTightness} we remark on several ways in which the assumptions are necessary and conclusions as strong as possible. Below, we give a complete proof for part \ref{itmCrossoverTheorem2Goods} because it is short and intuitive. The proof of part \ref{itmCrossoverTheoremManyGoods}, which involves tedious approximations and concentration bounds, is deferred to Appendix \ref{appCrossoverProof}.

\begin{proof}[Proof of Theorem \ref{thmCrossover} part \ref{itmCrossoverTheorem2Goods}.]
	
	By Lemma \ref{lemOneGoodNotDivided}, the divider will always leave one good undivided. Since the chooser's values for both goods are drawn independently from the same distribution $\mathcal{D}$, which is supported only over positive values, the chooser is more likely to pick the pile with the undivided good. Therefore, it must be that the divider's least-preferred good is the one that remains undivided, for otherwise he could have a higher expected utility switching the roles of the two goods. In other words, for some function $f$ (that depends on the fixed distribution $\mathcal{D}$), the optimal division is always as follows: given values $g^D_1$ and $g^D_2$, the divider places a $p^D := f(\{g^D_1, g^D_2\}) > \frac12$ fraction of good $\arg\max_i(g^D_i)$ in pile 1, with the rest of that good and all of the other good in pile 2. Analogously, let $p^C := f(\{g^C_1, g^C_2\})$, i.e., the amount of the chooser's preferred good that would have gone into pile 1 had the chooser been the divider.
	
	To compare the two players' ex-ante expected utilities, we fix realizations of $g^D_1$, $g^D_2$, $g^C_1$, and $g^C_2$, and compare the chooser's actual utility with the hypothetical utility had the roles been reversed. There are three possible cases to consider, depending on the realizations of $g^D_1$, $g^D_2$, $g^C_1$, and $g^C_2$.
	
	\textbf{Case 1:} The two players weakly prefer different goods. In this case, the chooser will receive all of her favorite good in pile 2. Had she instead been the divider, she would have only received a $p^C$-fraction of her favorite good. Since $p^C \leq 1$, the chooser is weakly better off.
	
	\textbf{Case 2:} The two players strictly prefer the same good, and $p^D \geq p^C$. In this case, the chooser receives a $p^D$-fraction of her favorite good in pile 1. As in Case 1, had she instead been the divider, then she would have only received a $p^C$-fraction of her favorite good. Since $p^C \leq p^D$, the chooser is weakly better off.
	
	\textbf{Case 3:} The two players strictly prefer the same good, and $p^D < p^C$. In this case, had the chooser been the divider, the divider would have opted for the chooser's preferred pile 1, since we know the divider would have weakly preferred a $p^D$-fraction of his favorite good to everything else, so he must have strictly preferred a $p^C$-fraction of his favorite good. Thus, if roles had been reversed, the chooser would have received her least-preferred pile, obtaining utility at most her proportionality guarantee. Since the chooser always can get at least this utility, the chooser is weakly better off being the chooser.
	
	In short, we have shown that the chooser is weakly better off in all cases. Furthermore, the chooser is strictly better off in Case 1 whenever $p^C < 1$. If this never occurs, that means that the divider always puts the goods entirely into different piles, in which case the divider receives utility equal to his proportionality guarantee and the chooser exceeds her proportionality guarantee (since the assumptions on $\mathcal{D}$ imply the chooser's values for the two goods are not always the same), so we are done. Assuming that $p^C < 1$ with nonzero probability, we further observe that, conditioning on $p^C < 1$, the players will prefer different goods (or be indifferent) at least half of the time. Thus, we see that the chooser is strictly better off with nonzero probability, so she is strictly better off overall.
\end{proof}

Theorem \ref{thmCrossover} appears to hold for normal priors with a positive mean as well (even though they may take unbounded and negative values). For $n = 2$, the chooser is better off, while for large $n$, the divider is better off. At what value of $n$ does the crossover occur? Figure \ref{figCrossoverNormal} plots estimated utilities per good for an empirical experiment where all values are drawn from $\nrm{1, 0.04}$. The crossover appears to be around 15 goods.

\ipnc{.8}{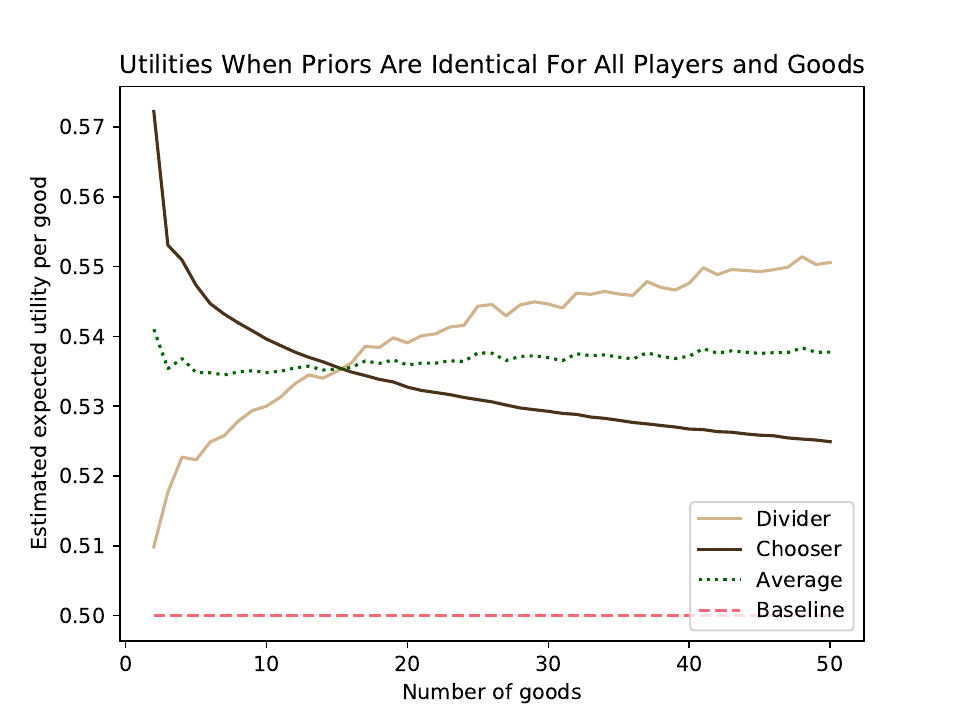}{\label{figCrossoverNormal}Estimated utilities from repeated trials of the following experiment with independent values. We first draw divider values $g^D_1, g^D_2, \dots, g^D_n$ i.i.d.\txt{} from $\nrm{1, 0.04}$, then compute the optimal division $p$ with respect to those values, with $\mathcal{G}^C_i = \nrm{1, 0.04}$ for each good $i$ as well. We compute the divider's and chooser's expected utility exactly with respect to $p$ and $g^D_1, g^D_2, \dots, g^D_n$, where the expectation is taken over the unknown values $g^C_1, g^C_2, \dots, g^C_n$. All utilities are averaged, then normalized by dividing by the number of goods $n$.}

\subsection{Heterogeneity of values}

Thus far we have been measuring and comparing the utilities afforded by the two roles ex-ante. However, there is more to the story: the realizations of the values will generally affect which role is more desirable. Figure \ref{figDeviationDeterminesRole} plots random samples of the values of 13 goods drawn from the same distribution, $\nrm{1, 0.04}$, as in Figure \ref{figCrossoverNormal}. We chose the number of goods to be 13 since that is near the crossover point in Figure \ref{figCrossoverNormal}, where the two roles have similar ex-ante utilities.\footnote{Even though the expected utilities are closer at 15 goods, the probability that the chooser is better off than the divider is closer to $\frac12$ at 13 goods than at 15 goods.} As one can see, when values vary more widely it is better to be the divider. This is because the divider's optimal strategy is to place all goods that he values highly in pile 1; he cares much less about the low-value goods. On the other hand, when values are more concentrated around the mean, it is better to be the chooser. In such a scenario, the divider's strategy will not yield a substantially high payoff. Instead, the chooser benefits simply from the fact that she will probably be able to take a larger and more valuable pile under the divider's optimal strategy.

\ipnc{.8}{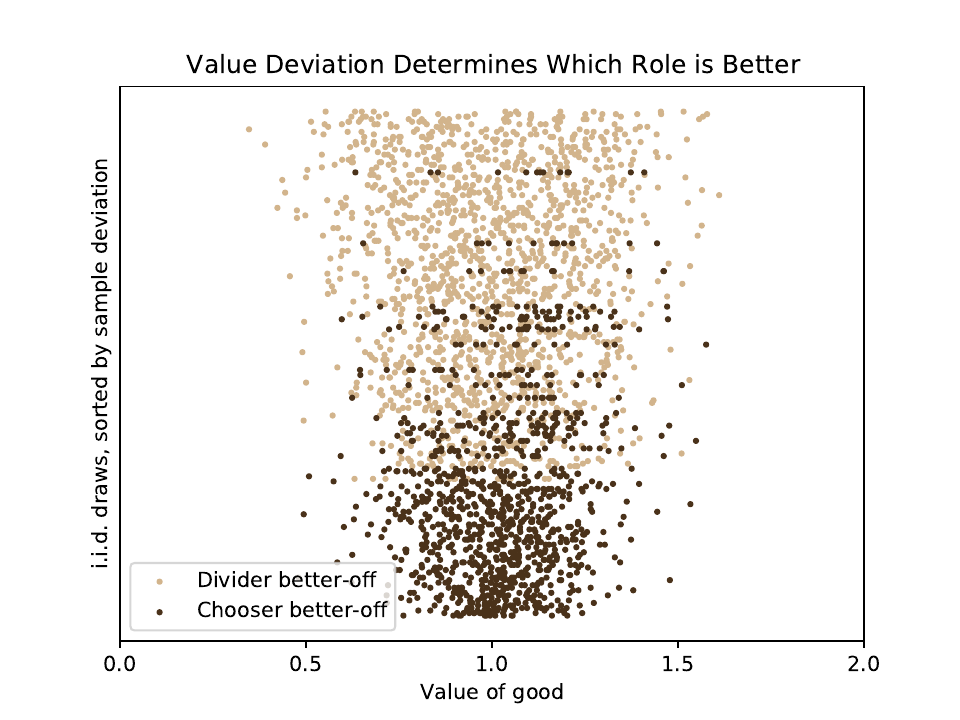}{\label{figDeviationDeterminesRole}200 random samples of values for 13 goods drawn i.i.d.\txt{} from $\nrm{1, 0.04}$, colored by the role that is better off given such values. Each row of points represents a separate draw, and the draws are sorted from bottom to top in increasing sample deviation (sum of absolute differences between the value of good $i$ and the mean of 1, across all 13 goods). Divider utilities are computed as the optimal utilities from Algorithm \ref{algNormalOpt}. Chooser utilities are estimated by averaging utilities with respect to a fixed ensemble of 4,000 optimal divisions under random divider values drawn from $\nrm{1, 0.04}$.}

\section{Conclusions}\label{secConclusion}

Divide-and-choose methods are widely employed by business partners going separate ways, inheritors dividing assets, and divorcing couples. They also lie at the heart of an array of managerial decisions. 

The divide-and-choose game is not unfamiliar to economics and computer science. However, typical formulations---such as the famed cake-cutting model---assume the players' preferences are known, or at the very least can be elicited truthfully. In the real world, by contrast, a divider will merely have subjective beliefs about the chooser’s preferences. This analysis pioneers the study of optimal strategies in divide-and-choose when one's counterpart's preferences are unknown.

A major finding is that the known and unknown preference situations lead to qualitatively different strategies. Moreover, a number of the results for the unknown preferences case are surprising. For example, with known preferences, the ratio of values for the two players, what we label the critical ratio, proves critical. All the goods in one pile have a higher ratio than any of the goods in the other. In addition, at most one good is divided between the piles, with division via a lottery if that good is physically indivisible. With unknown preferences, by contrast, assignment by critical ratios may be violated. That is, fractions allocated to pile 1 may be non-monotonic in the ratios. Moreover, it might be optimal to fractionally allocate up to $n - 1$ out of $n$ goods.

Uncertainty on preferences makes diversification a vital concern. It competes with efficient division as an objective for the divider. The analysis proceeds with a pile 1, preferred by the divider, with its complement denoted pile 2. The divider is eager to get the chooser to pick the latter. This is fostered in part by assigning greater chooser expected value to pile 2. This process is limited because the total value in pile 1 is being diminished, and that is the pile the divider hopes to get and will get most of the time. Holding the disparity in expected values fixed, the chooser is more likely to opt for pile 1 the greater is variability in actual values. Diversification---achieved by dividing multiple goods between the two piles---significantly reduces the variability in the chooser's valuation of piles. Hence, it reduces the likelihood that the divider ends up with pile 2.

Computing the divider’s optimal division is easy in the known-preferences case. It is surprisingly
difficult with preferences unknown. We already mentioned the troubling lack of monotonicity in critical ratios. Worse still, multiple local optima are commonly encountered. Despite these challenges, we are pleased to identify algorithms that compute divisions for normal and uniform priors in polynomial time, approximating the optimal utility to within arbitrary precision. We suspect that exactly computing optimal divisions in either of these settings is NP-Hard. Characterizing the precise computational complexity is an interesting open question.

We investigated some general effects of divider risk aversion. While this makes the divider's optimization task more computationally difficult, our results do shed some light on how to actually find optimal divisions. \emph{Chooser} risk aversion, not addressed here at all, merits future study. If goods are divisible, chooser risk aversion is irrelevant. With indivisible goods only divided via lotteries, it could be consequential. One intriguing question is whether correlating lotteries across multiple goods can increase the divider's expected utility.

Many of our results have direct implications for real-world divide-and-choose situations, implying that  the models from which they flow represent stylized facts.  For example, a typical buy-sell arrangement allows either player to grab the yet unclaimed divider’s role.  The relative attraction of the divider’s role is greater the more refined one’s assessment of the other player’s preferences, the better informed the other player about your preferences, and the greater the number of goods.  Both the divider and the chooser roles gain in value the more extreme one’s preferences relative to expectation; the parameters of the specific situation will determine which role is preferred.

Insights from divide-and-choose apply in a range of contexts that are not specifically divide-and-choose. A closely related setting occurs, for example, when one party makes a take-it-or-leave-it offer, such as a job offer, to another party. The recipient can accept the offer or stick with the status quo. Such offers commonly include divisible goods. Thus, the offer of a university to a job candidate might specify a salary and a teaching load, where there is a permissible range for each attribute.

There is an indirect equivalent to our formulation. An accepted offer, the university’s preferred outcome, is the equivalent of pile 1.  A rejected offer is the status quo. (No resources change hands given a rejection.) The chooser accepts when he prefers the offer to the status quo. In the traditional divide-and-choose situation, the divider is uncertain about the chooser’s valuation of the two piles. In a take-it-or-leave it offer, she is uncertain about the chooser’s valuation of the offer relative to the status quo. On a technical level, these two games are slightly different, where the latter game essentially involves the same optimization problem except that the $q_i$ variables must be nonnegative and $P$ can be greater than $\frac12$. Thus we would expect our analyses about critical ratios and diversification to still apply.

Posit that the university and the candidate have linear and additive preferences over the two attributes. However, the university has only priors on the candidate’s weights on attributes and valuation of the status quo. It would like to increase the probability of acceptance, i.e., that the offer lies above the candidate’s unknown reservation value, but also wants to increase its value from an accepted offer. An offer using interior values for some attributes---a direct analogue to fractional allocations in our model---will often be optimal. Interior values provide diversification; they increase the probability of acceptance for a given value of an accepted offer to the university.

As a final example, many bargaining contexts involve reciprocal concessions on different issues over time as a means to build confidence. Thus, a labor union might initially give up on its four-day-a-week demand in exchange for the corporation’s willingness to fund pensions more generously. When offering such preliminary trades, each party will reason about critical ratios to decide the optimal offer or counteroffer. But it will also be thinking ahead as to what it will have to give up on the big issue, say salary, to get an acceptance. While our model is formally situated as a one-shot interaction, we believe our insights extend qualitatively to these more complex negotiations: at every step on the path to a deal, the parties must think about both critical ratios and effective diversification.




There are intriguing areas for future work, only touched on in this paper, that will require a blend of analytic advances and attention to real-world features. They include player risk aversion, ex-post renegotiation of agreed-upon allocations, and mechanisms---such as alternating choices---that mix responsibilities for playing divider and chooser.

Of course, a salient feature of the real world is that players are not perfectly rational. An interesting theoretical question for future work would be to investigate an alternative equilibrium concept where players make mistakes with small probability. Also, an experimental study is currently underway to see how dividers and choosers actually play in a game closely resembling our formulation.

Allocation systems, in a range of areas from divide-and-choose to the market, involve players
acting strategically to maximize their take given what they know of the other player’s
valuation. This paper hopes to have shown that, in the allocation of analytic attention, the art of actually \emph{playing} the divide-and-choose game has received a smaller pile than it deserves.

\section*{Acknowledgments}
	We are deeply grateful to Jack Stade, who suggested the proof approach for Theorem \ref{thmSymmetricConvexity}; to Pranay Gorantla, who found and alerted us to the Meyer-Reisner Theorem (Theorem \ref{thmMR2}); and to Yannai Gonczarowski, for his careful reading of an earlier draft with extremely helpful feedback.
	
	We also thank Justin Chan for numerous suggestions that improved readability, as well as our anonymous reviewers at both Management Science and EC 2023. Finally, we thank Paul G\"olz, Gregory Kehne, Prashanth Amireddy, Elizabeth Pratt, and Santhoshini Velusamy, who also consulted on Theorem \ref{thmSymmetricConvexity}, and David Eppstein, who consulted on Theorem \ref{thmUniformOpt}.
	
	This material is based upon work supported by the National Science Foundation Graduate Research Fellowship Program under Grant No.\txt{} DGE1745303, and by the Mossavar-Rahmani Center for Business and Government, Harvard University. Any opinions, findings, and conclusions or recommendations expressed in this material reflect the views of the authors and not necessarily those of the National Science Foundation or the Mossavar-Rahmani Center.

\bibliographystyle{plain}
\bibliography{bibliography}

\appendix
	\section{Proof of Lemma \ref{lemBaselineUtilities}}\label{appBaselineUtilities}
	
	To prove \ref{itmBaselineChooser}, observe that the average utility from the chooser's two options is
	$$\frac12\left(\sum_{i = 1}^n p_i g^C_i + \sum_{i = 1}^n (1 - p_i) g^C_i\right) = \frac12 \sum_{i = 1}^n g^C_i,$$
	so at least one of the options must yield at least this utility. To prove \ref{itmBaselineDivider}, observe that the division $p = \left(\frac12, \frac12, \dots, \frac12\right)$ yields utility
	$\sum_{i = 1}^n \left(\frac12\right) g^D_i = \frac12 \sum_{i = 1}^n g^D_i$
	when the chooser picks pile 2 and utility
	$\sum_{i = 1}^n \left(1 - \frac12\right) g^D_i = \frac12 \sum_{i = 1}^n g^D_i$
	when the chooser picks pile 1. As the divider always has a strategy guaranteeing utility $\frac12 \sum_{i = 1}^n g^D_i$, at equilibrium he must receive at least that utility in expectation.
	\qed
	
	\section{Proof of Lemma \ref{lemStructureGeneral}}\label{appProofStructureGeneral}
	We first show that optimal divisions exist. Using Equation (\ref{equDefinePFromQ}), We may express the divider's expected utility as
	\begin{align}
		\ee[u^D] &= P \left(\sum_{i = 1}^{n} (1 - p_i) g^D_i\right) + (1 - P) \left(\sum_{i = 1}^{n} p_i g^D_i\right)\nonumber\\
		&= P \left(\sum_{i = 1}^{n} \left(1 - \frac{q_i}{2} - \frac12\right) g^D_i\right) + (1 - P) \left(\sum_{i = 1}^{n} \left(\frac{q_i}{2} + \frac12\right) g^D_i\right)\nonumber\\
		&= P \sum_{i = 1}^{n} \frac{g^D_i}{2} - P \sum_{i = 1}^{n} \frac{g^D_i q_i}{2} + \sum_{i = 1}^{n} \frac{g^D_i q_i}{2} + \sum_{i = 1}^{n} \frac{g^D_i}{2} - P \sum_{i = 1}^{n} \frac{g^D_i q_i}{2} - P \sum_{i = 1}^{n} \frac{g^D_i}{2}\nonumber\\
		&= \sum_{i = 1}^n \frac{g^D_i}{2} + \left(\frac12 - P\right) \sum_{i = 1}^n q_i g^D_i.\label{equFinalTerm}
	\end{align}
	The optimal utility is attained by maximizing Equation (\ref{equFinalTerm}) over the variables $\seq{q}{i} \in [-1, 1]$. Let $u^*$ denote the supremum of this optimal utility, and consider a sequence of divisions $q^1, q^2, q^3, \dots$ whose expected utilities converge to $u^*$. Let $q^* \in [-1, 1]$ be the division in the limit of a convergent subsequence. We claim that $q^*$ yields the optimal utility $u^*$. Indeed, since Equation (\ref{equFinalTerm}) is a continuous function of $q$ and $P$, this could only fail if there were a discontinuity in the function $P(q)$ at $q^*$. Such a discontinuity must be the result of a set of chooser types $S$ of positive probability mass where all types in $S$ pick the divider's strictly preferred pile at $q^*$ and the other pile after an arbitrarily small deviation from $q^*$. But this means that all chooser types in $S$ are indifferent between the two piles, which contradicts our assumption that an indifferent chooser breaks her indifference in favor of the divider. Thus, $q^*$ attains the optimal utility $u^*$.
	
	If a given division does not satisfy Equation (\ref{equDividerPrefersPile1Q}), then sending each $q_i \mapsto -q_i$ will satisfy it. This corresponds to sending $p_i \mapsto 1 - p_i$, so it is an equivalent division up to renaming the piles. Thus, it is without loss of generality to assume (\ref{equDividerPrefersPile1Q}) holds in an optimal division.
	
	If $P > \frac12$, then, since the final sum in Equation (\ref{equFinalTerm}) is nonnegative by (\ref{equDividerPrefersPile1Q}), we have
	$$\ee[u^D] \leq \sum_{i = 1}^n \frac{g^D_i}{2},$$
	so the divider is no better off than his proportionality guarantee. Therefore, the divider is at least equally well-off setting $p = \left(\frac12, \frac12, \dots, \frac12\right)$, in which case it is without loss of generality that $P \leq \frac12$.
	
	To prove the final statement, simply observe that the final term in Equation (\ref{equFinalTerm}) is nonzero if and only if the inequalities in both Equations (\ref{equDividerPrefersPile1Q})	and (\ref{equChooserProbablyPrefersPile2Q}) are strict.
	\qed
	
	\section{Proof of Lemma \ref{lemNormalAlgoCorrect}}\label{appProofNormalAlgoCorrect}
	Observe that each division $\seq{p}{n}$ computed by Algorithm \ref{algNormalOpt} is valid, since $0 \leq p_i \leq 1$ if and only if $-1 \leq q_i \leq 1$, which is enforced by the first constraint of $\mathcal{C}_P$. We claim that, on each iteration of the main loop, the division $\seq{p}{n}$ computed by Algorithm \ref{algNormalOpt} achieves an interim expected utility of $\ee[u^D] \geq u_P$. (In fact, the utility will be exactly $u_P$, but equality is not necessary to prove.)
	
	The chooser weakly prefers pile 1 if and only if
	$$\sum_{i = 1}^{n} g^C_i p_i \geq \sum_{i = 1}^{n} g^C_i (1 - p_i).$$
	Note that this is equivalent to
	$$s := \sum_{i = 1}^{n} g^C_i q_i \geq 0.$$
	Since each $g^C_i$ follows a normal distribution with mean $\mu_i$ and variance $\sigma_i^2$, we know that $s$ follows a normal distribution with mean
	$$\sum_{i = 1}^{n} \mu_i q_i$$
	and variance
	$$\sum_{i = 1}^{n} \sigma_i^2 q_i^2.$$
	Hence, the probability that $s \geq 0$ is given by
	$$1 - \Phi\left(\frac{0 - \sum_{i = 1}^{n}\mu_i q_i}{\sqrt{\sum_{i = 1}^{n} \sigma_i^2 q_i^2}}\right) = \Phi\left(\frac{\sum_{i = 1}^{n}\mu_i q_i}{\sqrt{\sum_{i = 1}^{n} \sigma_i^2 q_i^2}}\right).$$
	Since the algorithm computes optimal $\seq{q}{n}$ on each iteration of the main loop to satisfy the third constraint, we know that this probability is at most $P$. Therefore,
	\begin{align*}
		\ee[u^D] &= \Pr[\txt{chooser picks pile 1}] \sum_{i = 1}^{n} g^D_i (1 - p_i) + \Pr[\txt{chooser picks pile 2}] \sum_{i = 1}^{n} g^D_i p_i\\
		&= \sum_{i = 1}^{n} g^D_i p_i + \Pr[\txt{chooser picks pile 1}]\left(\sum_{i = 1}^{n} g^D_i (1 - p_i) - \sum_{i = 1}^{n} g^D_i p_i\right)\\
		&\geq \sum_{i = 1}^{n} g^D_i p_i + P\left(\sum_{i = 1}^{n} g^D_i (1 - p_i) - \sum_{i = 1}^{n} g^D_i p_i\right) \stextn{since the second constraint ensures the term in parentheses is nonpositive}\\
		&= \sum_{i = 1}^{n}g^D_i (P (1 - p_i) + (1 - P) p_i)\\
		&= \sum_{i = 1}^{n}g^D_i \left(P \left(1 - \left(\frac{q_i}{2} + \frac12\right)\right) + \left(1 - P\right) \left(\frac{q_i}{2} + \frac12\right)\right)\\
		&= \sum_{i = 1}^{n}\frac{g^D_i}{2} \left(P\left(1 - q_i\right) + (1 - P)\left(1 + q_i\right)\right)\\
		&= u_P.
	\end{align*}
	Thus, the claim is proved.
	
	Let $(p^*_1, p^*_2, \dots, p^*_n)$ denote an optimal division from Lemma \ref{lemStructureGeneral}, yielding interim expected utility $u^*$, and let $(q^*_1, q^*_2, \dots, q^*_n)$ denote the respective auxiliary variables for this division (i.e., obtained from Equation (\ref{equDefineQFromP})). Recall that, in this optimal division, the divider weakly prefers pile 1, and the probability that the chooser picks pile 1 is $P^* \leq \frac12$. Therefore, on some iteration of Algorithm \ref{algNormalOpt},
	$$P - \delta \leq P^* \leq P.$$
	Let $(\seq{q}{n})$ denote the optimal solution to $\mathcal{C}_P$ on this iteration, with optimal value $u_P$, and let $(\seq{p}{n})$ denote the corresponding division.
	
	Observe that $(q^*_1, q^*_2, \dots, q^*_n)$ is feasible for $\mathcal{C}_P$. To see this, note that the first constraint is satisfied by the fact that it corresponds to a valid division with each $p_i \in [0, 1]$. The second constraint is satisfied because we are assuming the divider prefers pile 1. Finally, for the third constraint, since $P^* \leq P$ implies $\Phi(P^*) \leq \Phi(P)$, we have
	$$\frac{\sum_{i = 1}^{n}\mu_i q^*_i}{\sqrt{\sum_{i = 1}^{n} \sigma_i^2(q^*_i)^2}} \leq \Phi(P^*) \leq \Phi(P).$$
	Thus, $(q^*_1, q^*_2, \dots, q^*_n)$ is a feasible solution for $\mathcal{C}_P$.
	
	Therefore, denoting the objective function of $\mathcal{C}_P$ by $f_P$, we have that the utility of the optimal solution returned by the algorithm is
	\begin{align*}
		\ee[u^D] &\geq u_P \stext{from the previous claim}\\
		&= f_P(\seq{q}{n})\\
		&\geq f_P(q^*_1, q^*_2, \dots, q^*_n) \stext{since $(q^*_1, q^*_2, \dots, q^*_n)$ is feasible and $(\seq{q}{n})$ is optimal}\\
		&= \sum\limits_{i = 1}^n\frac{g^D_i}{2} \left(P\left(1 - q^*_i\right) + (1 - P)\left(1 + q^*_i\right)\right)\\
		&= \sum_{i = 1}^n\frac{g^D_i}{2} \left(1 + q^*_i\right) - P\sum_{i = 1}^n g^D_i q^*_i\\
		&= \sum_{i = 1}^n\frac{g^D_i}{2} \left(1 + q^*_i\right) - P^*\sum_{i = 1}^n g^D_i q^*_i - (P - P^*)\sum_{i = 1}^n g^D_i q^*_i\\
		&= \sum_{i = 1}^n\frac{g^D_i}{2} \left(1 + 2p^*_i - 1\right) - P^*\sum_{i = 1}^n g^D_i (2p^*_i - 1) - (P - P^*)\sum_{i = 1}^n g^D_i q^*_i\\
		&= P^* \sum_{i = 1}^{n} g^D_i (1 - p^*_i) + (1 - P^*) \sum_{i = 1}^{n} g^D_i p^*_i - (P - P^*) \sum_{i = 1}^{n} g^D_i q^*_i\\
		&= \Pr[\txt{chooser picks 1}] \sum_{i = 1}^{n} g^D_i (1 - p^*_i) + \Pr[\txt{chooser picks 2}] \sum_{i = 1}^{n} g^D_i p^*_i - (P - P^*) \sum_{i = 1}^{n} g^D_i q^*_i\\
		&= u^* - (P - P^*) \sum_{i = 1}^{n} g^D_i q^*_i\\
		&\geq u^* - \abs{(P - P^*)} \sum_{i = 1}^{n} \abs{g^D_i} \abs{q^*_i}\\
		&\geq u^* - \delta \sum_{i = 1}^{n} \abs{g^D_i} \stext{since $P - \delta \leq P^*$ and each $\abs{q_i} \leq 1$}\\
		&= u^* - \gamma
	\end{align*}
	as desired. \qed
	
	\section{Proof of Theorem \ref{thmUniformOpt}}\label{appFPRAS}
	
	For any $0 \leq P \leq \frac12$ and $V > 0$, let
	$$S_{P, V} := \{q \in [-1, 1]^n \suchthat \Pr_{g^C \sim \overline{\mathcal{G}}^C}[q \cdot g^C \geq 0] \leq P \txt{ and } q \cdot g^D \geq V\}.$$
	This is the set of divisions for which the chooser picks Pile 1 with probability at most $P$ and Pile 1 is worth at least $V$ more to the divider than Pile 2. We will describe an algorithm, running in polynomial time in the size of the input, $\frac{1}{\varepsilon}$, and $\frac{1}{\delta}$, that either finds a division $q \in S_{P + \varepsilon, V}$ or reports that $S_{P, V}$ is empty, and is correct with probability $\geq 1 - \delta$. It is straightforward to see that such an algorithm can be called repeatedly to give the desired FPRAS: we apply the same search over $P$ as in Algorithm \ref{algNormalOpt}. However, instead of solving the quadratic program, we repeatedly test feasibility of $S_{P, V}$, using a binary-search to find the maximal $V$.
	
	Consider the following randomized separation oracle $\mathcal{O}$ for $S_{P, V}$. Given a queried point $q$, first we test whether $q \cdot g^D \geq V$. If not, then we return this inequality as a violated constraint. Otherwise, we empirically estimate the probability $r$ that the chooser picks Pile 1 by repeatedly sampling many chooser values $g^C$ and counting the number of times $q \cdot g^C \geq 0$. If the empirical estimate $\widehat{r}$ is at most $P + \frac{2\varepsilon}{3}$, then we return true. Otherwise, we compute an estimate $\widehat{m}$ of the centroid $m$ of the set
	$$M := \{g^C \in \prod_{i = 1}^n [a_i, b_i] \suchthat q \cdot g^C = 0\}$$
	and return the violated linear constraint that $q \cdot \widehat{m} < 0$. Computing the approximate centroid $\widehat{m}$ can be accomplished by repeatedly sampling many uniformly random points from $M$ and taking the average, which can be done in polynomial time using the hit-and-run random walk \cite{LVSampling}. We omit the tedious details concerning the sample complexity of estimating $r$ and $m$. All we require is that, for any $\delta' > 0$, we can, in polynomial time in $\frac{1}{\varepsilon \delta'}$, ensure with probability at least $1 - \delta'$ that $\widehat{r}$ is within $\frac{\varepsilon}{3}$ of the true probability $r$ and $\widehat{m}$ is within
	$$\frac{\varepsilon}{3}\min_{1 \leq i \leq n}(b_i - a_i)$$
	of the true centroid $m$ (under the $L_1$ norm). We say $\mathcal{O}$ \emph{misses} if one of these two conditions is false.
	
	We run the ellipsoid algorithm using separation oracle $\mathcal{O}$ until it returns true. This takes a number of iterations $s$ that is at most a polynomial function of the size of the input and $\frac{1}{\varepsilon}$. Letting $\delta' = \frac{\delta}{s}$, each call to $\mathcal{O}$ will take polynomial time, and we know by the union bound that the probability some iteration misses is at most $\delta$. Thus, to show that the algorithm is correct, we just need to show that $\mathcal{O}$ returns a correct result whenever it does not miss. There are three cases to consider, depending on what $\mathcal{O}$ returns when queried on division $q$.
	
	If $\mathcal{O}(q)$ returns true, then $q \cdot g^D \leq V$ and $\widehat{r} \leq P + \frac{2\varepsilon}{3}$. Assuming $\mathcal{O}$ does not miss, we will have $\abs{\widehat{r} - r} \leq \frac{\varepsilon}{3}$, in which case $\widehat{r} \leq P + \frac{2\varepsilon}{3} + \frac{\varepsilon}{3} = P + \varepsilon$ by the triangle inequality. Hence, the algorithm has found a division $q \in S_{P + \varepsilon, V}$ as desired.
	
	If $\mathcal{O}(q)$ returns that $q$ violates the inequality $q \cdot g^D \geq V$, then it is always correct (in fact, there was no way for it to miss). It has determined that $q$ does not satisfy this constraint, and obviously any $q' \in S_{P, V}$ does satisfy it.
	
	Finally, consider the case where $\mathcal{O}(q)$ returns that $q$ violates the inequality $q \cdot \widehat{m} < 0$. Observe that $q$ does indeed violate this inequality because $\widehat{m} \in M$. All that remains to show is that all $q' \in S_{P, V}$ do satisfy it. We show the contrapositive, that an arbitrary $q'$ violating the inequality is \emph{not} in $S_{P, V}$. So suppose $q' \cdot \widehat{m} \geq 0$. Without loss of generality, assume that $q_n$ has the largest magnitude out of all $q_i$ for all $1 \leq i \leq n$. Note that $q_n \neq 0$, for otherwise $q$ is the all-zero vector, so $q \cdot g^D \geq V$ would have been violated, as we are assuming $V > 0$. Since $q' \cdot \widehat{m}$ is positive, we have that
	\begin{align*}
		\Pr_{g^C \sim \overline{\mathcal{G}}^C}[q' \cdot g^C \geq 0] &\geq \Pr_{g^C \sim \overline{\mathcal{G}}^C}[q' \cdot g^C - q' \cdot \widehat{m} \geq 0]\\
		&= \Pr_{g^C \sim \overline{\mathcal{G}}^C}[q' \cdot g^C \geq q' \cdot \widehat{m}]\\
		&\geq \Pr_{g^C \sim \overline{\mathcal{G}}^C}[q' \cdot g^C \geq q' \cdot m] - \Pr_{g^C \sim \overline{\mathcal{G}}^C}[q' \cdot \widehat{m} \geq q' \cdot g^C \geq q' \cdot m]\\
		&= \Pr_{g^C \sim \overline{\mathcal{G}}^C}[q' \cdot (g^C - m) \geq 0] - \Pr_{g^C \sim \overline{\mathcal{G}}^C}\left[g^C_n \in J\right] \txt{ where} \push J := \left[\frac{1}{q'_n} \left(\sum_{i = 1}^{n}q'_i \widehat{m}_i - \sum_{i = 1}^{n - 1}q'_i g^C_i\right), \frac{1}{q'_n} \left(\sum_{i = 1}^{n}q'_i m_i - \sum_{i = 1}^{n - 1}q'_i g^C_i\right)\right]\\
		&> \left(P + \frac{\varepsilon}{3}\right) - \frac{\varepsilon}{3} \stext{see justification below}\\
		&= P.
	\end{align*}
	Therefore, $q' \notin S_{P, V}$. It remains to justify the final inequality above, which follows from the following two inequalities:
	\begin{align}
		\Pr_{g^C \sim \overline{\mathcal{G}}^C}[q' \cdot (g^C - m) \geq 0] &> P + \frac{\varepsilon}{3} \label{equUniformOpt1}\\
		\Pr_{g^C \sim \overline{\mathcal{G}}^C}\left[g^C_n \in J\right] &\leq \frac{\varepsilon}{3} \label{equUniformOpt2}
	\end{align}
	
	To prove Inequality (\ref{equUniformOpt1}), let $H_1$ and $H_2$ be the $(n - 1)$-dimensional planes
	\begin{align*}
		H_1 := \{g^C \in \rr^n \suchthat q' \cdot (g^C - m) = 0\}, && H_2 := \{g^C \in \rr^n \suchthat q \cdot g^C = 0\},
	\end{align*}
	with positive sides defined as
	\begin{align*}
		H_1^+ := \{g^C \in \rr^n \suchthat q' \cdot (g^C - m) \geq 0\}, && H_2^+ := \{g^C \in \rr^n \suchthat q \cdot g^C \geq 0\}.
	\end{align*}
	Recall that $r$ is the true probability that $q \cdot g^C \geq 0$, and $\widehat{r}$ is the estimate of $r$. Applying Theorem \ref{thmMR2} part \ref{itmMR2Centroid} to $H_1$, we know that $m \in K_{r}$. Since $q' \cdot (m - m) = 0$, we have $m \in H_2^+$  as well, so $m \in H_2^+ \cap K_r$. Hence, applying Theorem \ref{thmMR2} part \ref{itmMR2Iff} (backward direction) to $H_2$, we derive that
	\begin{equation*}
		\Pr_{g^C \sim \overline{\mathcal{G}}^C}[q' \cdot (g^C - m) \geq 0] = \Pr_{g^C \sim \overline{\mathcal{G}}^C}[g^C \in H_2^+] \geq r.
	\end{equation*}
	Assuming $\mathcal{O}$ did not miss, we additionally know that $\abs{r - \widehat{r}} \leq \frac{\varepsilon}{3}$ and $\widehat{r} > P + \frac{2\varepsilon}{3}$, so Inequality (\ref{equUniformOpt1}) follows from the triangle inequality.
	
	Finally, we prove Inequality (\ref{equUniformOpt2}) holds under any fixed values of $g^C_i$ for $1 \leq i \leq n - 1$, letting $g^C_n$ be the only random variable. Using the assumptions that $q'_n$ has the largest magnitude and that $\mathcal{O}$ did not miss, we know that
	$$\abs{J} = \sum_{i = 1}^n \frac{q'_i}{q'_n}(\widehat{m}_i - m_i) \leq \sum_{i = 1}^n \abs{\widehat{m}_i - m_i} \leq \frac{\varepsilon}{3}(b_n - a_n).$$
	Therefore, since $g^C_n \sim [a_n, b_n]$, the probability $g^C_i \in J$ is at most $\frac{\varepsilon}{3}$, so Inequality (\ref{equUniformOpt2}) holds.
	
	This proves that the separation oracle is correct with probability $1 - \delta$, and thus so is the algorithm. \qed
	
	\section{Proof of Lemma \ref{lemNormalAlgoCorrect}}\label{appProofDiscreteAlgoCorrect}
	Suppose $\overline{\mathcal{G}}^C$ is an arbitrary distribution supported on a finite set $\{\seq{x}{\ell}\} \subseteq \rr^n$. For each $1 \leq j \leq \ell$, we write $r_j$ for the probability that $g^C = x_j$. The main idea for finding an optimal division is similar to that of Algorithm \ref{algNormalOpt}: we try to guess the set of chooser types who pick pile 1; we then use linear programming to compute the optimal division with respect to the additional constraints that entails. (Since there are an exponential number of subsets in contention, this means the algorithm is only practical when the number of types $\ell$ is small.) Specifically, consider the following algorithm.\\
	
	\begin{algorithm}[H]
		\SetAlgoLined
		\DontPrintSemicolon
		\KwIn{Divider values $g^D_1, g^D_2, \dots, g^D_n$ (not all zero), possible chooser value vectors $\seq{x}{\ell} \in \rr^n$, and corresponding probabilities $\seq{r}{\ell}$.}
		\KwOut{An optimal division $\seq{p}{n}$.}
		$u \gets -\infty$\;
		\For{$S \subseteq [\ell]$}{
			$P \gets \sum_{j \in S} r_j$\;
			\If{$P \leq \frac12$}{
				$u', \seq{q}{i} \gets$ optimal solution to the following linear program $\mathcal{C}_S$:
				$$\lpmax{u' = \sum\limits_{i = 1}^n\frac{g^D_i}{2} \left(P\left(1 - q_i\right) + (1 - P)\left(1 + q_i\right)\right)}{\\$-1 \leq q_i \leq 1$ & for all $1 \leq i \leq n$,\\$\sum\limits_{i = 1}^{n} (x_j)_i q_i \leq 0$ & for all $j \in [\ell] \setminus S$}$$
				\If{$u' > u$}{
					$u \gets u'$\;
					\For{$i \gets 1, 2, \dots, n$}{
						$p_i \gets \frac{q_i + 1}{2}$\;
					}
				}
			}
		}
		\Return{$(\seq{p}{n})$}\;
		\caption{Computes an optimal division given the divider's values and an arbitrary discrete prior for the chooser's values.}
		\label{algDiscreteOpt}
	\end{algorithm}
	$ $
	
	Let $p^* = (p^*_1, p^*_2, \dots, p^*_n)$ denote an optimal division from Lemma \ref{lemStructureGeneral}, with divider utility $u^*$ and probability $P^*$ that the chooser picks pile 1. Let $S^* \subseteq [\ell]$ denote the set of chooser types $j$ such that, when the chooser has value vector $g^C = x_j$, she chooses pile 1 given the division $p^*$. Note that the probability the chooser picks pile 1 given $p^*$ is $P^* = \sum_{j \in S^*} r_j \leq \frac12$. It is not too hard to see that $\mathcal{C}_S$ finds the optimal division given the additional constraint that all types $j \notin S$ pick pile 2, assuming further that all types $j \in S$ will pick pile 1 (which can only lower the objective value). In particular, this means that $u'$ is always at least the utility from some feasible division, and on the iteration of the main loop when $S = S^*$, Algorithm \ref{algDiscreteOpt} will successfully find $p = p^*$, and the optimal utility will be $u = u^*$. The claim about running time follows from the observation that each $\mathcal{C}_S$ is a linear program. \qed
	
	\section{Proof of Lemma \ref{lemOneGoodNotDivided}}\label{appOneGoodNotDivided}
	
	Let $q$ be an optimal division from Lemma \ref{lemStructureGeneral}. If $q$ achieves a higher-than-proportional utility, then there must be some good $i$ such that $\abs{q_i} > 0$. Let $i^*$ be a good maximizing $\abs{q_{i^*}}$. If $\abs{q_{i^*}} = 1$, then good $i^*$ is put entirely into one of the two piles; otherwise, we claim that the divider can improve the division by linearly scaling $q$, dividing each component by $\abs{q_{i^*}}$. Since
	$$P = \Pr\left[\sum_{i = 1}^n q_i g^C_i > 0 \right] = \Pr\left[\sum_{i = 1}^n \frac{q_i}{\abs{q_{i^*}}} g^C_i > 0 \right],$$
	the chooser is just as likely to pick pile 1, but now the divider obtains a greater utility from pile 1, so he receives a higher expected utility.
	\qed
	
	\section{Tightness of Theorem \ref{thmCrossover}}\label{appCrossoverTightness}
	
	In this section we provide counterexamples to possible ways that one might hope to strengthen/generalize Theorem \ref{thmCrossover}.
	
	\begin{itemize}
		\item \textbf{The condition on $\mathcal{D}$ that the critical ratios are not always identical.} Otherwise the divider may not be able to beat his proportionality guarantee. For instance, in the extreme case where the values of the two goods are always identical, both roles confer the same utility, which would violate both parts of the theorem.
		\item \textbf{The assumption that values are not correlated for $n = 2$.} Suppose $\mathcal{D}$ is a distribution where the divider faces no uncertainty of the chooser's value after seeing his own value. Then the divider can extract all of the surplus from having different values, leaving the chooser with utility equal to her proportionality guarantee. Hence, correlations can make the divider better off, even for only two goods.
		\item \textbf{Positivity of $\mathcal{D}$.} If $\mathcal{D}$ is the uniform distribution on $[-1, 1]^2$,  one can see that $P = \frac12$ in any division, so the divider cannot surpass his proportionality guarantee utility for any number of goods. However, for large $n$, the chooser will certainly reap more than her proportional utility (assuming the divider breaks his indifference between divisions in a way that benefits the chooser).
		\item \textbf{The condition in Equation (\ref{equCrossoverSubexpTailAssumption}).} Roughly, this says that the values from $\mathcal{D}$ are typically not extremely small. To see why this assumption is necessary, suppose the divider's value for each good is 0 with probability 90\% and 1 with probability 10\%, while the chooser's value is uniformly distributed on $[0, 1]$. For a large number of goods, the divider will place the roughly 10\% of the goods for which he has high-value in pile 1 and the remaining roughly 90\% of the goods in pile 2. By choosing pile 2, the chooser obtains an ex-ante expected utility of $0.9 \cdot 0.5 = 0.45$ per good. If she were the divider, she would have to create piles of roughly even size (with the preferred pile 1 containing the best half of the goods, slightly smaller than pile 2), only receiving an ex-ante utility of $0.5 \cdot 0.75 = 0.375$ per good. Thus, in this example, it is favorable for the first player to be the chooser. However, this is only because the second player sometimes draws the value zero. As we show in the proof, as long as the divider's value is not zero, even the slightest bit greater than zero, the optimal division leaves the chooser with only her proportionality utility guarantee.
		
		\item \textbf{The chooser is better off with two goods but not always with three goods.} Suppose each of the three goods have value 1 with 99\% probability and and 300 with 1\% probability. There are three cases to consider.
		\begin{itemize}
			\item With probability $(\frac{99}{100})^6 \approx 0.941$, all six valuations will be the low value of 1. The divider will divide the goods evenly and the chooser will pick an arbitrary pile, with both players gaining the same utility.
			\item With probability $6(\frac{99}{100})^5(\frac{1}{100}) \approx 0.057$, there will be exactly one high-valued good. If it is the divider's value, the divider will optimally place it alone, entirely in pile 1 and most of the other two goods in pile 2. The divider's expected utility is at least $(\frac{99}{100}) \cdot 300 = 297$ and the chooser's expected utility is at most 2. If the chooser has the high valuation, the divider will evenly divide the goods between the piles, so there is a chance the chooser's favorite good will be fractionally allocated. Specifically, some good must be divided with at least $\frac{1}{4}$ in each pile. So the chooser's expected utility from the high-value good is at most $(\frac{2}{3})\cdot300 + (\frac{1}{3})\cdot(\frac{3}{4})\cdot300 = 275$, for a total utility of at most 278. Thus, in this case, the expected utility of the divider is higher than that of the chooser by at least
			$$\frac12(297 - 2) + \frac12(0-278) = 8.5$$
			\item With probability $15(\frac{99}{100})^4(\frac{1}{100})^2 \approx 0.00144$, there will be exactly two high-valued goods. We can upper bound the chooser's utility in this case by the expected value of all goods to the chooser, which is $300 + 1 + 1 = 302$.
			\item With the remaining probability of $1 - (\frac{99}{100})^6 - 6(\frac{99}{100})^5(\frac{1}{100}) - 15(\frac{99}{100})^4(\frac{1}{100})^2 \approx 0.00002$, more than two goods have high value, in which case we may upper bound the chooser's utility by the maximal value of the sum of all goods, which is 900.
		\end{itemize}
		In total, the expected utility of the divider minus that of the chooser is at least
		$$0.057 \cdot 8.5 - 0.00144 \cdot 302 - 0.00002 \cdot 900 \approx 0.03 \geq 0.$$
	\end{itemize}
	
	\section{Proof of Theorem \ref{thmCrossover} part \ref{itmCrossoverTheoremManyGoods}}\label{appCrossoverProof}
	
	Assume without loss of generality that the support of $\mathcal{D}$ is contained in $[0, 1] \times [0, 1]$. Note that this means the maximum possible utility for either player is at most $n$. Let
	$$B := \frac{1}{2} \ee_{\substack{(g^1, g^2) \sim \mathcal{D}}}\left[g^1\right].$$
	We will show that, for all sufficiently large $n$, the player whose values are drawn according to the first component of $\mathcal{D}$ attains an expected utility of at least $n(B + \varepsilon)$ as the divider for some constant $\varepsilon > 0$, but attains utility strictly less than this as the chooser for any $\varepsilon > 0$.
	
	We begin with the divider's utility. Since the critical ratio of a good is not always the same value, there must be a pair of values $b_1 > b_2 > 0$, such that the critical ratio of a good is at least $b_1$ with some nonzero probability, and at most $b_2$ with some nonzero probability. By similar reasoning, we can write $b_1 = \frac{c_1}{c_3}$ such that, with nonzero probability, the numerator of the critical ratio is bounded below by $c_1$ and the denominator is bounded above by $c_3$; likewise, we can write $b_2 = \frac{c_2}{c_4}$ such that, with nonzero probability, the numerator of the critical ratio is bounded above by $c_2$ and the denominator is bounded below by $c_4$. Note that $b_1 > b_2$ implies
	\begin{equation}\label{equFourConstants}
		c_1 c_4 > c_2 c_3.
	\end{equation}
	By independence of the values of goods $i$ and $i + 1$, it follows that, with some probability $c_5 > 0$, good $i$ will have a higher critical ratio than good $i + 1$ as according to these bounds:
	\begin{equation}\label{equIInS}
		\Pr_{\substack{g^D_i, g^C_i \sim \mathcal{D}\\g^D_{i + 1}, g^C_{i + 1} \sim \mathcal{D}}}[g^D_i \geq c_1,\ g^D_{i + 1} \leq c_2,\ \ee[g^C_i] \leq c_3,\ \txt{and } \ee[g^C_{i + 1}] \geq c_4] \geq c_5.
	\end{equation}
	We let $S$ be the set of odd $i < n$ for which the condition in Equation \ref{equIInS} is satisfied. Consider the following division, which is reminiscent of the proof of Proposition \ref{proStructureNormal}:
	\begin{align*}
		q_i := \threecases{\txt{if } i \in S}{\frac{c_2 + c_4}{2\alpha}}{\txt{if }i - 1 \in S}{-\frac{c_1 + c_3}{2\alpha}}{\txt{otherwise }}{0} && \txt{where} && \alpha := \max\left\{\frac{c_1 + c_3}{2}, \frac{c_2 + c_4}{2}\right\}
	\end{align*}
	
	The divider strictly prefers pile 1 by
	\begin{align*}
		\sum_{i = 1}^n q_i g^D_i &= \sum_{i \in S}\left(\frac{c_2 + c_4}{2\alpha} g^D_i - \frac{c_1 + c_3}{2\alpha} g^D_{i + 1}\right)\\
		&\geq \sum_{i \in S} \left(\frac{1}{2\alpha} (c_2 + c_4) c_1 - \frac{1}{2\alpha} (c_1 + c_3) c_2\right)\\
		&= \frac{\abs{S}}{2\alpha} (c_1 c_4 - c_2 c_3),
	\end{align*}
	which is a constant greater than zero. Likewise, the interim expectation of the difference between the two piles' value for the chooser is
	\begin{align*}
		\ee\left[\sum_{i = 1}^n q_i g^C_i\right] &= \sum_{i \in S} \left(\frac{c_2 + c_4}{2\alpha} \ee[g^C_i] - \frac{c_1 + c_3}{2\alpha} \ee[g^C_{i + 1}]\right)\\
		&\leq \sum_{i \in S} \left(\frac{c_2 + c_4}{2\alpha} c_3 - \frac{c_1 + c_3}{2\alpha} c_4\right)\\
		&= \frac{\abs{S}}{2\alpha} (c_2 c_3 - c_1 c_4),
	\end{align*}
	which is a constant less than zero. Thus, by standard concentration bounds, for large enough $n$ the probability the chooser picks pile 1 converges to zero, so the divider's utility gains beyond the proportionality guarantee converges to
	$$\frac{\ee[\abs{S}]}{4 \alpha} (c_1 c_4 - c_2 c_3) = \frac{n c_5(c_1 c_4 - c_2 c_3)}{4 \alpha}.$$
	Thus, letting $\varepsilon := \frac{c_5(c_1 c_4 - c_2 c_3)}{5 \alpha} > 0$, the desired conclusion follows.
	
	We now compute the utility of the first component player if they are the chooser. We need to be slightly more formal in stating the concentration bounds required for this case. Specifically, given $\varepsilon > 0$, we choose $n$ large enough so that $n \geq \frac{4}{\varepsilon}$ and
	\begin{equation*}\label{equChooseN}
		\Pr_{(g^1, g^2) \sim \mathcal{D}}\left[g^2 \leq 8n \exp\left(-\frac18 \varepsilon^2 n\right)\right] < \frac{\varepsilon}{4n}.
	\end{equation*}
	Note that this is a restatement of Equation (\ref{equCrossoverSubexpTailAssumption}) applied to $\frac{\varepsilon}{4}$. It follows from the union bound that
	\begin{equation}\label{equChooseNUnionBound}
		\Pr_{(g^1, g^2) \sim \mathcal{D}^n}\left[\min_{i \in [n]}\{g^2_i\} \leq 8n \exp\left(-\frac18 \varepsilon^2 n\right)\right] < \frac{\varepsilon}{4}.
	\end{equation}
	
	For any division $q \in [-1, 1]^n$, let $X = X(q)$ be the random variable as defined in Equation (\ref{equDefineX}), which is the chooser's value for pile 1 minus the chooser's value for pile 2. Write $\mu_X \in [-n, n]$ for the expectation of $X$ conditioned on the divider's values. Assume we draw a list of divider values $g^2_1, g^2_2, g^2_3, \dots, g^2_n$ such that the event in Equation (\ref{equChooseNUnionBound}) does \emph{not} occur, i.e.,
	\begin{equation}\label{equEventDoesNotOccur}
		\min_{i \in [n]}\{g^2_i\} > 8n \exp\left(-\frac18 \varepsilon^2 n\right)
	\end{equation}
	Then, under any division,
	\begin{align*}
		\Pr\left[\abs{X - \mu_X} \geq \frac{\varepsilon}{2}n\right] &\leq \Pr\left[\abs{X - \mu_X} \geq \frac{\varepsilon}{2}n - 1\right]\\
		&\leq \Pr\left[\abs{X - \mu_X} \geq \frac{\varepsilon}{4}n\right] \snc{n \geq 4/\varepsilon}\\
		&\leq 2\exp\left(-\frac{2(\frac{\varepsilon}{4}n)^2}{\sum_{i = 0}^n(1 - 0)^2}\right) \stext{Hoeffding's inequality}\\
		&= 2 \exp\left(-\frac18 \varepsilon^2 n\right)\\
		&< \frac{\min_{i \in [n]}\{g^2_i\}}{4n} \stext{from Equation (\ref{equEventDoesNotOccur})}\\
		&\leq \frac{1}{4n}\\
		&\leq \frac{1}{n}\\
		&\leq \frac{\varepsilon}{4} \snc{n \geq 4/\varepsilon}.
	\end{align*}
	There are three inequalities derived from this chain that we will use shortly:
	\begin{align}
		\Pr\left[\abs{X - \mu_X} \geq \frac{\varepsilon}{2}n\right] &< \frac{1}{4n} < \frac12\label{equHoeffding0}\\
		\Pr\left[\abs{X - \mu_X} \geq \frac{\varepsilon}{2}n - 1\right] &< \frac{\min_{i \in [n]}\{g^2_i\}}{4n} \label{equHoeffding1}\\
		\Pr\left[\abs{X - \mu_X} \geq \frac{\varepsilon}{2}n\right] &< \frac{\varepsilon}{4}. \label{equHoeffding2}\\
	\end{align}
	
	We claim that, still assuming Equation (\ref{equEventDoesNotOccur}) holds, $\abs{\mu_X} \leq \frac{\varepsilon}{2} n$ under any optimal division. First suppose that $\mu_X > \frac{\varepsilon}{2}n$. By Equation (\ref{equHoeffding0}), with probability at least $\frac12$, we will have $\abs{X - \mu_X} < \frac{\varepsilon}{2} n$, in which case the triangle inequality implies $\abs{X} > 0$. This means the chooser will pick pile 1 with probability strictly greater than $\frac12$, which cannot happen in an optimal division, so we have a contradiction. Next suppose that $\mu_X < -\frac{\varepsilon}{2}n$. Note that any optimal division $q$ has some $q_j \leq 0$; otherwise the chooser picks pile 1 with certainty. Take such a $j$, and consider the alternative division $q'$ where $q'_j := q_j + 1$ and $q'_i := q_i$ for all $i \neq j$. Let $X'$ be the new difference in pile utilities under $q'$, with mean $\mu_{X'}$. Let $P$ and $P'$ be the respective probabilities that the chooser picks pile 1 in $q$ and $q'$, and let $u$ and $u'$ be the respective expected divider utilities. Since the value of the $j\tth$ good has expectation at most 1, we have $\mu_{X'} \leq \mu_{X'} + 1 < -(\frac{\varepsilon}{2}n - 1)$. Thus, $X' > 0$ implies $\abs{X' - \mu_{X'}} \geq \frac{\varepsilon}{2}n - 1$, so
	\begin{equation}\label{equPPrimeBound}
		P' = \Pr[X' > 0] \leq \Pr\left[\abs{X' - \mu_{X'}} \geq \frac{\varepsilon}{2}n - 1\right] < \frac{\min_{i \in [n]}\{g^2_i\}}{4n} \leq \frac{(q_j' - q_j)g^2_j}{4n}.
	\end{equation}
	In the middle inequality, we have applied Equation (\ref{equHoeffding1}) to $X'$, which is valid since the bound holds under any division. Note in particular this means that $P' \leq \frac14$. We may thus lower bound the utility gain of the divider from switching to $q'$ as follows:
	\begin{align*}
		u' - u &= \left(\sum_{i = 1}^n \frac{g^2_i}{2} + \left(\frac12 - P'\right) \sum_{i = 1}^n q'_i g^2_i\right) - \left(\sum_{i = 1}^n \frac{g^2_i}{2} + \left(\frac12 - P\right) \sum_{i = 1}^n q_i g^2_i\right) \stextn{by Equation (\ref{equFinalTerm})}\\
		&= \left(\frac12 - P'\right) \sum_{i = 1}^n q'_i g^2_i - \left(\frac12 - P\right) \sum_{i = 1}^n q_i g^2_i\\
		&= \left(\frac12 - P'\right) \left(\sum_{i = 1}^n q'_i g^2_i -  \sum_{i = 1}^n q_i g^2_i\right) + \left(\left(\frac12 - P'\right) - \left(\frac12 - P\right)\right) \sum_{i = 1}^n q_i g^2_i\\
		&= \left(\frac12 - P'\right) (q'_j - q_j) g^2_j + \left(P - P'\right) \sum_{i = 1}^n q_i g^2_i\\
		&\geq \left(\frac12 - P'\right) (q'_j - q_j) g^2_j - P' \sum_{i = 1}^n q_i g^2_i\\
		&> \left(\frac14\right) (q'_j - q_j) g^2_j - P' n \snc{P' < \frac14 \txt{ and each } q_i g^2_i \leq 1}\\
		&> \left(\frac14\right) P' \cdot 4n - P' n \stext{from Equation (\ref{equPPrimeBound})}\\
		&= 0.
	\end{align*}
	Thus, we have shown that $q'$ is a strictly better division, contradicting the assumption that $q$ was optimal. It follows that $\abs{\mu_X} \leq \frac{\varepsilon}{2}$.
	
	Therefore, we know from Equations (\ref{equChooseNUnionBound}) and (\ref{equHoeffding2}) that the conditions $\abs{X - \mu_X} < \frac{\varepsilon}{2} n$ and $\abs{\mu_X} \leq \frac{\varepsilon}{2}$ are each violated with probability less than $\frac{\varepsilon}{4}$. By the union bound, it follows that, with probability at least $1 - \frac{\varepsilon}{2}$, neither condition is violated. In this case, we know that $\abs{X} < \varepsilon n$ by the triangle inequality. Thus, with probability at least $1 - \frac{\varepsilon}{2}$, we may bound $X$ by $\varepsilon n$, whereas in the other case we may still bound $X$ by $n$. Putting these bounds together, we conclude that the chooser's expected utility is
	$$\ee[u^C] = nB + \frac{\ee[\abs{X}]}{2} \leq nB + \frac{\left(1 - \frac{\varepsilon}{2}\right) \varepsilon n + \left(\frac{\varepsilon}{2}\right)n}{2} < n(B + \varepsilon)$$
	as desired.
	\qed
\end{document}